\documentclass{jaa}
\usepackage{natbib} 
\usepackage{hyperref}
\usepackage{amsmath}
\usepackage{amstext}
\usepackage{amssymb}
\usepackage{url}
\usepackage{txfonts}
\usepackage{balance}
\usepackage{textcase}
\usepackage{float}
\usepackage{rotating,longtable}
\usepackage{xspace}
\usepackage{hyperref}
\usepackage{graphicx}
\usepackage{xcolor}

\usepackage{graphicx}
\usepackage{caption}
\usepackage{subcaption}
\usepackage{lipsum} % For dummy text only

\usepackage{lineno}

\usepackage[T1]{fontenc}
\usepackage{ae,aecompl}

\begin{document}\sloppy

\title{Probing Accretion and Outflow in V1180 Cas through High-Resolution Optical Spectroscopy}

\author{Tarak Chand\textsuperscript{1,2,*},Saurabh Sharma\textsuperscript{1}, Koshvendra Singh\textsuperscript{1}, Joe P. Ninan\textsuperscript{3}, Arpan Ghosh\textsuperscript{4}, Devendra K. Ojha\textsuperscript{3}, Jyotirmoy Dey\textsuperscript{3}, Manojit Chakraborty\textsuperscript{1,5}, Kartik Gokhe\textsuperscript{1}, Aayushi Verma\textsuperscript{1,2}, Mamta\textsuperscript{1,5} \and Ajay Kumar Singh\textsuperscript{6} }

% \author{Tarak Chand\textsuperscript{1,2,*}\orcidlink{0009-0008-8490-8601},
%  Saurabh Sharma\textsuperscript{1}\orcidlink{0000-0001-5731-3057}, Koshvendra Singh\textsuperscript{1}\orcidlink{0000-0002-7434-9681}, Joe P. Ninan\textsuperscript{3}\orcidlink{0000-0001-8720-5612}, Arpan Ghosh\textsuperscript{4}\orcidlink{0000-0001-7650-1870}, Devendra K. Ojha\textsuperscript{3}\orcidlink{0000-0001-9312-3816}, Jyotirmoy Dey\textsuperscript{3}\orcidlink{0000-0003-4074-4365}, Manojit Chakraborty\textsuperscript{1,5}\orcidlink{0009-0009-6420-8058}, Kartik Gokhe\textsuperscript{1}\orcidlink{0009-0004-4776-6780}, Aayushi Verma\textsuperscript{1,2}\orcidlink{0000-0002-6586-936X}, Mamta\textsuperscript{1,5}\orcidlink{0009-0001-4144-2281} \and Ajay Kumar Singh\textsuperscript{6}\orcidlink{0000-0001-7881-7748}
% }
\affilOne{\textsuperscript{1}Aryabhatta Research Institute of Observational Sciences (ARIES) Manora Peak, Nainital-263001, India.\\}
\affilTwo{\textsuperscript{2}M.J.P. Rohilkhand University Bareilly-243006, India.\\}
 \affilThree{\textsuperscript{3}Department of Astronomy and Astrophysics Tata Institute of Fundamental Research (TIFR) Mumbai-400005, India.\\}
 \affilFour{\textsuperscript{4}Instituto de Radioastronom{\'i}a y Astrof{\'i}sica, Universidad Nacional Aut{\'o}noma de M{\'e}xico, Antigua Carretera a P{\'a}tzcuaro 8701, Ex-Hda. San Jos{\'e} de la Huerta, Morelia, Michoac{\'a}n, M{\'e}xico C.P. 58089\\}
\affilFive{\textsuperscript{5}Indian Institute of Technology (IIT) Roorkee, Roorkee-247667, India.\\} \affilSix{\textsuperscript{6}Department of Applied Physics/Physics Bareilly College, Bareilly-243006, India. \\}
%%escape two column mode for title, affiliation and abstract
%%by giving \twocolumn command as shown

\twocolumn[{
\maketitle

%%include \corres to print the corresponding author Email id
\corres{tarakchands@gmail.com}
%\corres{saurabh175@gmail.com, saurabh@aries.res.in}

%%include \msinfo for
%%manuscript information such as
%%received, revised and accepted dates
%%
\msinfo{--}{--}

%%abstract
\begin{abstract}

We present an analysis of a high-resolution optical spectrum of V1180 Cas obtained with HIRES at the W. M. Keck Observatory during a bright photometric state of the source. The spectrum is dominated by strong emission lines, including H$\alpha$, the Ca II infrared triplet, He I, and O I, along with numerous Fe I and Fe II transitions and several forbidden lines such as [O I] and [S II], indicating ongoing accretion and mass-loss activity. A weak Li I $\lambda$6708 absorption feature confirms the youth of the source. Using the Li I absorption and selected Fe I emission lines, we report for the first time a radial velocity of $-16 \pm 3$ km s$^{-1}$ for V1180 Cas.
The H$\alpha$ and H$\beta$ profiles exhibit asymmetric structures with blueshifted absorption components, suggesting outflowing material along the line of sight. The He I $\lambda$5876 line displays a narrow component likely associated with post-shock accretion regions and a broad, slightly blueshifted component probably arising from magnetospheric flows and/or inner disk winds. The [O I] $\lambda$6300 profile is decomposed into low- and high-velocity components, tracing a slow disk wind and a fast jet. Using the forbidden [O I] line, we estimate for the first time a disk inclination angle of $\approx50^\circ$. The derived mass accretion and jet mass-loss rates imply $\dot{M}_{jet}/\dot{M}_{acc} \sim 0.01$--$0.03$, at the lower end of, but consistent with, the range observed for Class II YSOs. Forbidden-line diagnostics indicate densities $\sim10^{3}$ and $10^{6}$ cm$^{-3}$ with temperatures of $\sim10^{4}$ K, supporting a multi-component outflow scenario. Overall, the results support a picture in which V1180 Cas is an actively accreting, moderately inclined system hosting a multi-component outflow.

\end{abstract}

%%insert keywords separated by 3 hyphens using \keywords{words}
\keywords{ stars: formation, stars: pre-main-sequence, stars: outflows, stars: variables: T Tauri, techniques: spectroscopy, stars: individual: (V1180 Cas)\\}}]

%%close the twocolumn escape here

%%include \doinum{number}for the DOI number in the header
%%include \volnum{number} for the volume number in the header
%%include \year{yyyy} for  year of publication in the header
%%include \pgrange{num--num} page range of article in the header
%%include \artcitid{num} for the article citation id
%%include \lp to print last page of the article
%%include \setcounter{page}{pagenum} for the exact starting page of the article

\doinum{12.3456/s78910-011-012-3}
\artcitid{\#\#\#\#}
\volnum{000}
\year{2026}
\pgrange{1--}
\setcounter{page}{1}
\lp{1}

%%%%%%%%%%%%%%%%% BODY OF PAPER %%%%%%%%%%%%%%%%%%

%%%%%%%%%%%%%%%%% INTRODUCTION %%%%%%%%%%%%%%%%%%%%%%%%%%%%%%%%%%

\section{Introduction}
Young stellar objects (YSOs) display a wide range of spectroscopic signatures associated with ongoing accretion and mass-loss activity. High-resolution optical spectroscopy is particularly valuable for disentangling these processes, as it resolves individual emission and absorption features and can separate their kinematic components. In this wavelength range, hydrogen lines and the Ca II infrared triplet (IRT) are widely used tracers of magnetospheric accretion, while permitted lines such as He I probe the immediate star–disk interaction region \citep{Hartmann_2016ARA&A..54..135H}. At the same time, low-excitation forbidden transitions such as [O I], [S II], [N II], and [Fe II] provide key diagnostics of jets, winds, and shocked gas \citep{Pascucci_2023ASPC..534..567P}.

Class II YSOs with circumstellar accretion disks \citep{Lada_1987IAUS..115....1L} exhibit a broad range of accretion and outflow properties. Sources with relatively high accretion rates ($\gtrsim10^{-8}$ $M_{\odot}$ yr$^{-1}$) are more commonly associated with spatially resolved microjets \citep[e.g.,][]{Ray_2007prpl.conf..231R,Nisini_2018A&A...609A..87N}. In sources with more modest accretion rates ($\sim10^{-11}$--$10^{-8}$ $M_{\odot}$ yr$^{-1}$), outflows often remain spatially unresolved and are instead inferred from spectroscopic diagnostics, including blueshifted forbidden emission lines and P Cygni absorption in strong permitted transitions such as He I $\lambda10830$.
Among optical forbidden transitions, [O I] $\lambda6300$ is the brightest and most commonly detected line, appearing in $\sim$80--90\% of accreting T Tauri stars. It has therefore become one of the principal tracers of unresolved outflows in nearby star-forming regions. The [O I] $\lambda6300$ line profile frequently consists of two distinct kinematic components: a high-velocity component (HVC), typically associated with collimated jets and velocities of $\sim50$--300 km s$^{-1}$, and a low-velocity component (LVC), generally linked to slower disk winds with velocities often below 10 km s$^{-1}$ \citep{Hartigan_1995ApJ...452..736H, Kwan_1995ApJ...454..382K, Cabrit_1999A&A...343L..61C, Nisini_2018A&A...609A..87N, Banzatti_2019ApJ...870...76B}. The HVC of [O I] $\lambda6300$ is preferentially observed in sources with higher stellar masses and stronger accretion. Its luminosity correlates with the accretion rate, with a typical mass-loss to accretion ratio of $\sim0.1$ \citep{Ellerbroek_2013A&A...551A...5E, Nisini_2018A&A...609A..87N, 2025A&A...700A..32R}. In contrast, the LVC is detected in nearly all systems showing [O I] emission and spans the full range of T Tauri accretion luminosities \citep{Natta_2014A&A...569A...5N, McGinnis_2018A&A...620A..87M, Nisini_2018A&A...609A..87N}. Spatially and spectrally resolved observations further suggest that the LVC originates in slower, wider-angle disk winds \citep{Simon_2016ApJ...831..169S, Banzatti_2019ApJ...870...76B}.

V1180 Cas is an eruptive Class II YSO located in the Lynds 1340 molecular cloud and was first identified as a YSO by \citet{Kun_1994A&A...292..249K}. Long-term photometric monitoring has revealed large-amplitude variability, reaching $\sim6$ mag in the $I$ band, together with a characteristic blueing effect during deep fading events, attributed to an increased contribution from scattered light \citep{Kun_2011, Mutafov_2022RAA....22l5014M}. \citet{Kun_2011} also identified a wide-separation, optically invisible companion at a projected distance of $\sim3600$ au and reported an accretion rate of $\gtrsim1.6\times10^{-7}$ $M_{\odot}$ yr$^{-1}$ during a bright state ($I\sim15$ mag), which decreased by about an order of magnitude during a subsequent fading episode. Spectroscopic studies have further established the presence of both accretion and outflow activity. \citet{Antoniucci_2014A&A...565L...7A} reported a median accretion rate of $\sim3\times10^{-8}$ $M_{\odot}$ yr$^{-1}$ and mass-loss rates of $\sim4\times10^{-9}$ and $\sim4\times10^{-10}$ $M_{\odot}$ yr$^{-1}$ from atomic forbidden and H$_2$ emission lines, respectively. They also identified a chain of knots indicative of jet activity.

More recently, \citet{Chand_2026ApJ...998..204C} combined long-term photometry with multi-epoch optical to near-infrared spectroscopy and found that the variability of V1180 Cas likely has a hybrid origin, involving both variable circumstellar extinction and changes in accretion activity. Their observations yielded accretion rates of $\sim10^{-8}$--$10^{-7}$ $M_{\odot}$ yr$^{-1}$ during different brightness states, while persistent forbidden-line emission provided evidence for ongoing outflow activity. These results establish V1180 Cas as a particularly useful laboratory for investigating the relationship between accretion variability and mass-loss in an eruptive Class II YSO.

In this study, we present an analysis of a high-resolution optical spectrum of V1180 Cas with the aim of constraining its kinematic properties and characterizing its accretion and outflow diagnostics. 
Unlike previous low-resolution studies \citep[e.g.,][]{Chand_2026ApJ...998..204C}, the present HIRES spectrum allows us to resolve the kinematic structure of the accretion and outflow tracers, derive the systemic radial velocity, constrain the disk inclination, and decompose the [O I] $\lambda$6300 and He I $\lambda$5876 emission profiles into distinct velocity components.

The paper is structured as follows: Section \ref{sec:obs} describes the observations and data reduction. Section \ref{sec:res} presents the results of the spectral analysis together with comparisons to previous work on this and related sources. Section \ref{sec:disc} interprets these results and discusses their implications for the nature of V1180 Cas. Finally, Section \ref{sec:summary} summarizes the results.

%%%%%%%%%%%%%%%%%%%%%%% OBSERVATION %%%%%%%%%%%%%%%%%%%%%%%%%%%%%%%%%%%%%%%%%%%%%%%%%%

\section{Data}
\label{sec:obs}
The high-resolution optical spectrum of V1180 Cas was retrieved from the Keck Observatory Archive\footnote{\url{https://koa.ipac.caltech.edu/cgi-bin/KOA/nph-KOAlogin}}. The observation was obtained with the High Resolution Echelle Spectrometer (HIRES; \citealt{1993AAS...183.1101V}) at the W.M. Keck Observatory on UT 2016-08-18 in the Engineering program E297X titled "TCSU". HIRES was used with the Science Mosaic detector, consisting of three CCDs. At the time of the observation, V1180 Cas was in a bright ($I\approx15$ mag) and relatively stable photometric state (see Figure 2 of \citealt{Chand_2026ApJ...998..204C}). The observation was obtained with a $1200$ s exposure through a 0$^{\prime\prime}$.574 slit, providing a spectral resolution of $R\sim$$72{,}000$, corresponding to a velocity resolution of $\sim$4 km s$^{-1}$. The spectrum covers approximately 4800 to 9250 \r{A}. The seeing, estimated from the full width at half maximum (FWHM) of the object trace in the raw 2D spectrum along the cross-dispersion direction, is $\sim0^{\prime\prime}.6$.

The data reduction and wavelength calibration were performed using the MAuna Kea Echelle Extraction (MAKEE\footnote{\url{https://sites.astro.caltech.edu/~tb/makee/}}) HIRES data reduction package. The reduction includes tracing the echelle orders, defining the object and background regions, and extracting the one-dimensional spectrum. The wavelength calibration was performed using a ThAr arc lamp exposure with a root mean square (rms) $\sim 0.07$ \r{A}. The MAKEE pipeline also applied the heliocentric correction to the wavelength scale; for this observation, the correction was $\sim17$ km s$^{-1}$. The retrieved spectrum is not absolutely flux calibrated.

%%%%%%%%%%%%%%%%%%%%%%%%%%% Result and Analysis %%%%%%%%%%%%%%%%%%%%%%%%%%%%%%%%%%%%%%%%%%%%%%%

\begin{figure*}
    \centering
    \includegraphics[width=1\linewidth]{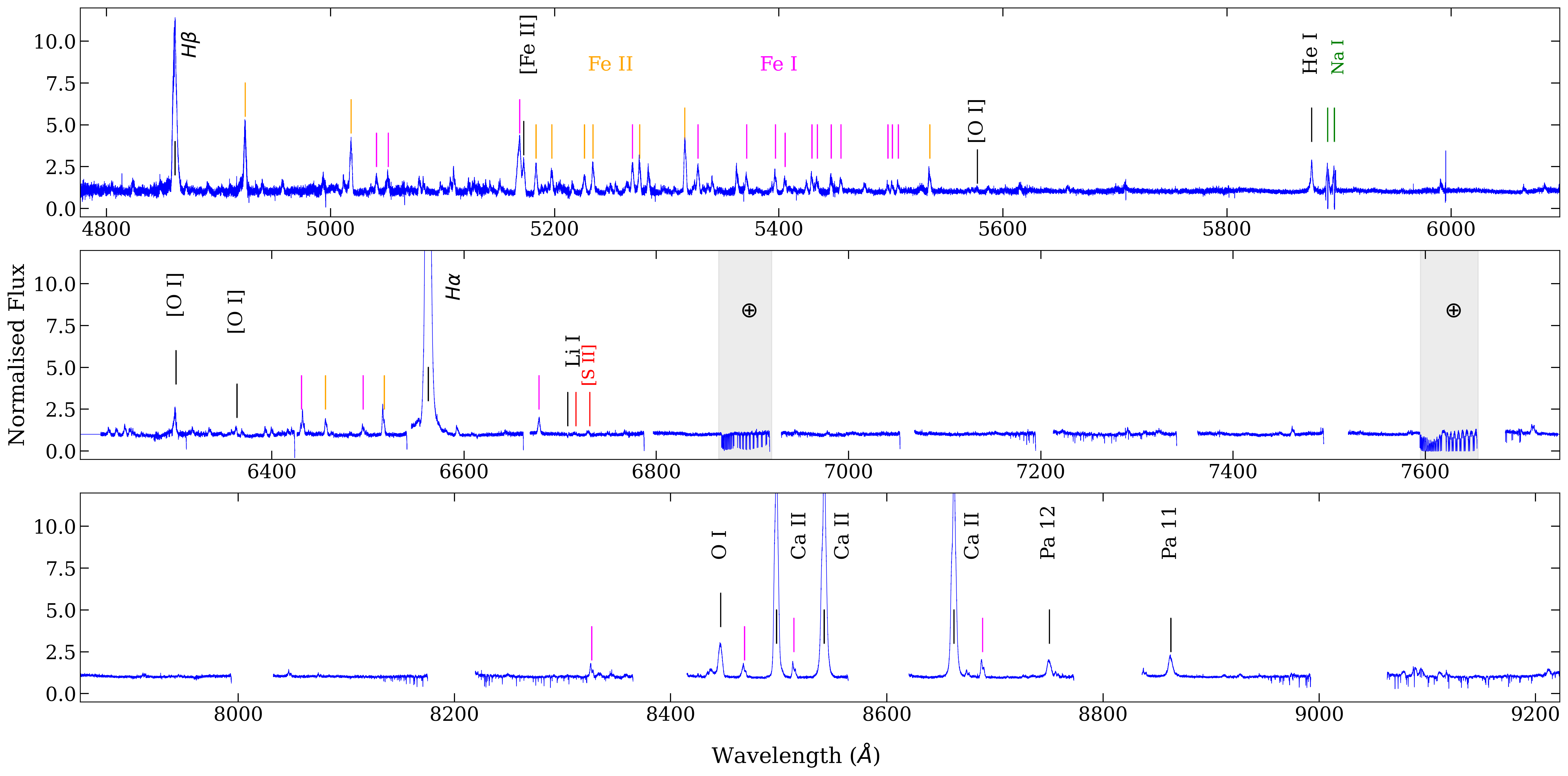}
    \centering
    \caption{High-resolution spectrum of V1180 Cas, normalized to the continuum level. Prominent spectral features are identified and labeled. The magenta and orange markers indicate Fe I and Fe II lines, respectively. Missing data points in the middle and lower panels correspond to gaps in wavelength coverage between adjacent echelle orders.}
    \label{fig:spec}
\end{figure*}

\section{Results and Analysis}
\label{sec:res}

\subsection{Spectral Features} \label{subsec:spec_feat}

The HIRES optical spectrum of V1180 Cas is shown in Figure \ref{fig:spec}. It is characterized by prominent emission in H$\alpha$ ($\lambda$6563), the Ca II IRT ($\lambda$8498, $\lambda$8542, and $\lambda$8662), Paschen 12 and 11 (Pa12 and Pa11; $\lambda$8750 and $\lambda$8863), O I $\lambda$8446, He I $\lambda$5876, and numerous Fe I and Fe II lines. In addition to these permitted transitions, a few forbidden lines are detected, including [O I] $\lambda$6300 and $\lambda$6364, [Fe II] $\lambda$5172, together with faint emission from [O I] $\lambda$5577 and [S II] $\lambda$6717 and $\lambda$6731. The Na I D lines at $\lambda$5890 and $\lambda$5896 are also present, exhibiting complex profiles with a U-shaped central absorption component. In addition to these emission features, a weak Li I $\lambda$6708 absorption feature is also detected. A brief description of these lines is given below:

\subsubsection{H$\alpha$ and H$\beta$ lines:}
The profiles of these lines are shown in the upper-left panel of Figure \ref{fig:line_profile}. Both lines exhibit asymmetric profiles with excess emission on the blue side, while their peak intensities are centered near $\sim$0 km s$^{-1}$. A shallow absorption feature is also present in the blue wing of each line. The blueshifted absorption components reach minimum intensity at velocities of approximately $-400$ and $-250$ km s$^{-1}$ for H$\alpha$ and H$\beta$, respectively.

Previously, \citet{Kun_2011} reported a weak P-Cygni absorption component in H$\alpha$ during one epoch of observation (2007 September). \citet{Chand_2026ApJ...998..204C} did not detect any P-Cygni absorption in H$\alpha$ during their decade-long monitoring campaign starting from 2015 August. This non-detection may be attributed to the lower spectral resolution ($R \sim 100$--1600) of their data.
%compared to the HIRES spectrum analyzed here.

\subsubsection{Ca II IRT lines:}
These lines display profiles broadly similar to those of H$\alpha$ and H$\beta$, with clear asymmetry, enhanced emission on the blue side, and peak intensities centered near $\sim$0 km s$^{-1}$ (see upper right panel in Figure \ref{fig:line_profile}). The measured intensity ratio of the Ca II IRT lines is approximately 1.0:1.1:1.1.

This ratio is similar to those previously reported for classical T Tauri stars \citep{Herbig_1980ApJ...242..628H, Hamann_1992ApJS...82..247H}, EXors \citep{Hodapp_2019AJ....158..241H, Hodapp_2020AJ....160..164H}, V1647 Ori \citep{Ninan_2013ApJ...778..116N}, and V899 Mon \citep{Ninan_2015ApJ...815....4N, Park_2021ApJ...923..171P}.

\begin{table*}
\centering
\centering
\caption{Detected Fe I lines in the spectrum. $E_i$, $g_i$, and $f_{ik}$ are the energy of the lower level, statistical weight, and oscillator strength of the transition, respectively. The relative optical depth values ($\tau_{rel}$) are normalized to the maximum derived $\tau_{rel}$. }
\begin{tabular}{ccccc}
\hline
\hline
        Line Wavelength& $E_i$ & $g_i$ & $f_{ik}$ & $\tau_{rel}^{norm}$ \\
              (\r{A})  &($eV$)& & &\\
        \hline
        Fe I 5041.071 & 0.96 & 7 & 0.000117 & 0.014 \\
        Fe I 5051.634 & 0.91 & 9 & 0.000178 & 0.031 \\
        Fe I 5168.897 & 0.05 & 7 & 0.000015 & 0.010 \\
        Fe I 5269.537 & 0.86 & 11 & 0.004340 & 1.000 \\
        Fe I 5328.038 & 0.91 & 9 & 0.003800 & 0.655 \\
        Fe I 5371.489 & 0.96 & 7 & 0.003240 & 0.397 \\
        Fe I 5397.127 & 0.91 & 9 & 0.001130 & 0.195 \\
        Fe I 5405.774 & 0.99 & 5 & 0.002860 & 0.237 \\
        Fe I 5429.696 & 0.96 & 7 & 0.001890 & 0.232 \\
        Fe I 5434.523 & 1.01 & 3 & 0.002520 & 0.121 \\
        Fe I 5446.916 & 0.99 & 5 & 0.002440 & 0.203 \\
        Fe I 5455.609 & 1.01 & 3 & 0.002700 & 0.130 \\
        Fe I 5497.516 & 1.01 & 3 & 0.000472 & 0.023 \\
        Fe I 5501.465 & 0.96 & 7 & 0.000128 & 0.016 \\
        Fe I 5506.779 & 0.99 & 5 & 0.000319 & 0.026 \\
        Fe I 6430.845 & 2.12 & 7 & 0.001410 & 0.022 \\
        Fe I 6677.985 & 2.69 & 11 & 0.003470 & 0.030 \\
        Fe I 8327.055 & 2.20 & 5 & 0.005970 & 0.057 \\
        Fe I 8468.406 & 2.22 & 3 & 0.002820 & 0.016 \\
        Fe I 8514.071 & 2.20 & 5 & 0.001180 & 0.011 \\
        Fe I 8688.625 & 2.18 & 7 & 0.008770 & 0.122 \\
\hline
\end{tabular}
\label{tab:FeI_info}
\end{table*}

\subsubsection{Fe I lines:}
The middle-left panel of Figure \ref{fig:line_profile} shows line profiles of the detected Fe I emission lines across the optical wavelength range. Most of these lines exhibit clear double-peaked structures with noticeable asymmetry, characterized by peak velocities at approximately $-35$ km s$^{-1}$ and $+45$ km s$^{-1}$ in the blue- and redshifted components, respectively.

To test whether this asymmetry could arise from self-absorption, we estimated the relative optical depths of the detected Fe I transitions. In general, the optical depth of an atomic transition depends on the physical conditions of the line-forming gas, including its temperature, density, chemical composition, and, to a lesser extent, microturbulence \citep{Calvet_1993ApJ...402..623C}. We assume that the observed Fe I lines originate from the same emitting region and therefore experience similar physical conditions. Under this assumption, the relative optical depths of different transitions can be approximated from the dependence of the line absorption coefficient on the population of the lower energy level and the oscillator strength of the transition. Using Equations 11.6 and 11.10 of \cite{2005oasp.book.....G}, for a bound--bound transition from a lower level $i$ to an upper level $k$, the line optical depth can be written as
\begin{equation}\label{eq:1}
\tau
= N_i \frac{\pi e^2}{m_e c} f_{ik} ~~\rm{or} ~~ \tau_{ik} \propto N_i f_{ik}
\end{equation}
where $N_i$ is the column density of Fe I atoms in the lower level and $f_{ik}$ is the oscillator strength. 
Under the assumption of local thermodynamic equilibrium (LTE), the population of the lower level follows the Boltzmann distribution,
\begin{equation}\label{eq:2}
N_i \propto N g_i
\exp\left(-\frac{E_i}{kT}\right),
\end{equation}
where $N$ is the total Fe I column density, $g_i$ is the statistical weight of the lower level, $E_i$ is its excitation energy, $k$ is the Boltzmann constant, and $T$ is the excitation temperature. Therefore, for transitions originating from the same region, for which $N$ and $T$ are assumed to be approximately common. Using the \ref{eq:1} and \ref{eq:2} relative optical depth can be expressed as %\citep{Beristain_1998ApJ...499..828B, Sicilia_2020A&A...643A..29S}
\begin{equation}
\label{eq:tau}
\tau_{\rm rel}
\propto
g_i f_{ik}
\exp\left(-\frac{E_i}{kT}\right).
\end{equation}
Thus, Equation \ref{eq:tau} does not provide an absolute measurement of the optical depth; rather, it provides a relative optical-depth proxy that allows us to compare the expected opacities of the different Fe I transitions under the adopted common physical conditions. We adopt $T=6500$ K as a representative temperature for this comparison. This choice is motivated by previous studies of Fe I and Fe II line-forming regions. \citet{Beristain_1998ApJ...499..828B} analyzed Fe I and Fe II emission in DR Tau and identified narrow and broad kinematic components, with the narrow component corresponding to temperatures of $\sim8000$--$14,000$ K and the broad component associated with a lower temperature, with an illustrative value of $\sim6900$ K. Similarly, \citet{Sicilia_2020A&A...643A..29S} found that Fe I/Ca I disk emission in Z CMa is associated with regions spanning approximately $2000$--$7000$ K, while Fe II traces comparatively higher-temperature regions. The adopted value of $6500$ K is also close to the characteristic temperature of $\sim6500$ K inferred by \citet{Sicilia_2012A&A...544A..93S} for the optical line-forming region of FU Ori from Fe I/Fe II line ratios. However, the latter analysis was based on absorption lines, whereas the Fe I lines considered here are observed in emission, and these may trace different physical conditions. Therefore, the adopted $6500$ K should not be interpreted as a temperature measurement for V1180 Cas, but only as a representative assumption for the present comparison. The atomic parameters ($g_i$, $E_i$, and $f_{ik}$) in Equation \ref{eq:tau} are taken from the NIST Atomic Spectra Database\footnote{\url{https://www.nist.gov/pml/atomic-spectra-database}}. The derived $\tau$ values, along with the corresponding atomic parameters, are listed in Table \ref{tab:FeI_info}. Interestingly, the relatively symmetric line at 5269.537 \r{A} exhibits the highest $\tau$ value, while several lines with lower $\tau$ values display pronounced asymmetry (e.g., $\lambda$8468, $\lambda$8514).

Multiple components in Fe I emission profiles have previously been reported in YSOs, and have been interpreted as arising from different regions of the accretion flow, with the narrow component (NC) associated with gas closer to the stellar surface and the broad component (BC) potentially originating near the inner disk edge \citep{Beristain_1998ApJ...499..828B, Sicilia_2012A&A...544A..93S, 2024A&A...690A.225A}.

The double-peaked, asymmetric Fe I profiles observed in V1180 Cas resemble the Fe I absorption-line profiles predicted by the accelerating disk wind models of FU Orionis presented by \citet[][see their Figure 7]{Calvet_1993ApJ...402..623C}. Although the Fe I lines in FU Orionis are observed in absorption rather than emission, the similarity in profile morphology may indicate comparable underlying kinematic structures.

\subsubsection{Forbidden lines:}
We detect a few forbidden emission lines in the spectrum of V1180 Cas, including [O I] $\lambda6300$, $\lambda6364$, $\lambda5577$, [S II] $\lambda6717$, $\lambda6731$, and [Fe II] $\lambda5172$. Among these, [O I] $\lambda6300$ is the strongest and most clearly resolved feature, while [Fe II] $\lambda5172$ is also prominent but partially blended with nearby Fe I emission. The [O I] $\lambda5577$ and [S II] $\lambda6717$, $\lambda6731$ lines are comparatively weak (see the middle-right panel of Figure \ref{fig:line_profile}). All detected forbidden lines are blueshifted and exhibit emission extending up to $\sim -150$ km s$^{-1}$.

Given the limited signal-to-noise ratio of the weaker transitions, we focus our detailed analysis on the [O I] $\lambda6300$ line (see Section \ref{subsec:line_prof_OI}). This line exhibits a multi-component structure consistent with the HVCs and LVCs commonly reported in YSOs \citep{Hartigan_1995ApJ...452..736H, Banzatti_2019ApJ...870...76B}.

\subsubsection{He I line:}
\begin{figure*}
    \centering
    \includegraphics[width=0.49\linewidth]{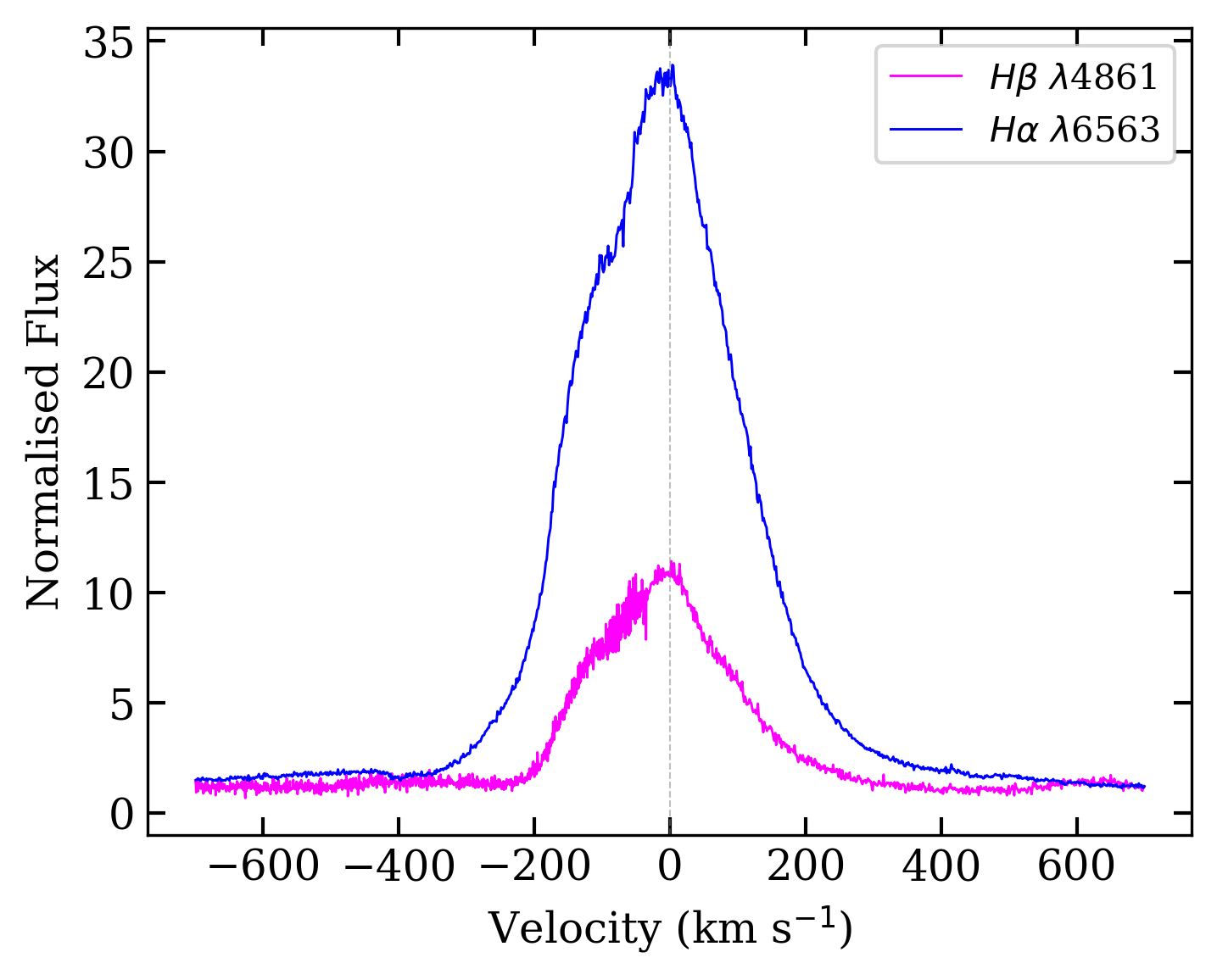}
    \includegraphics[width=0.49\linewidth]{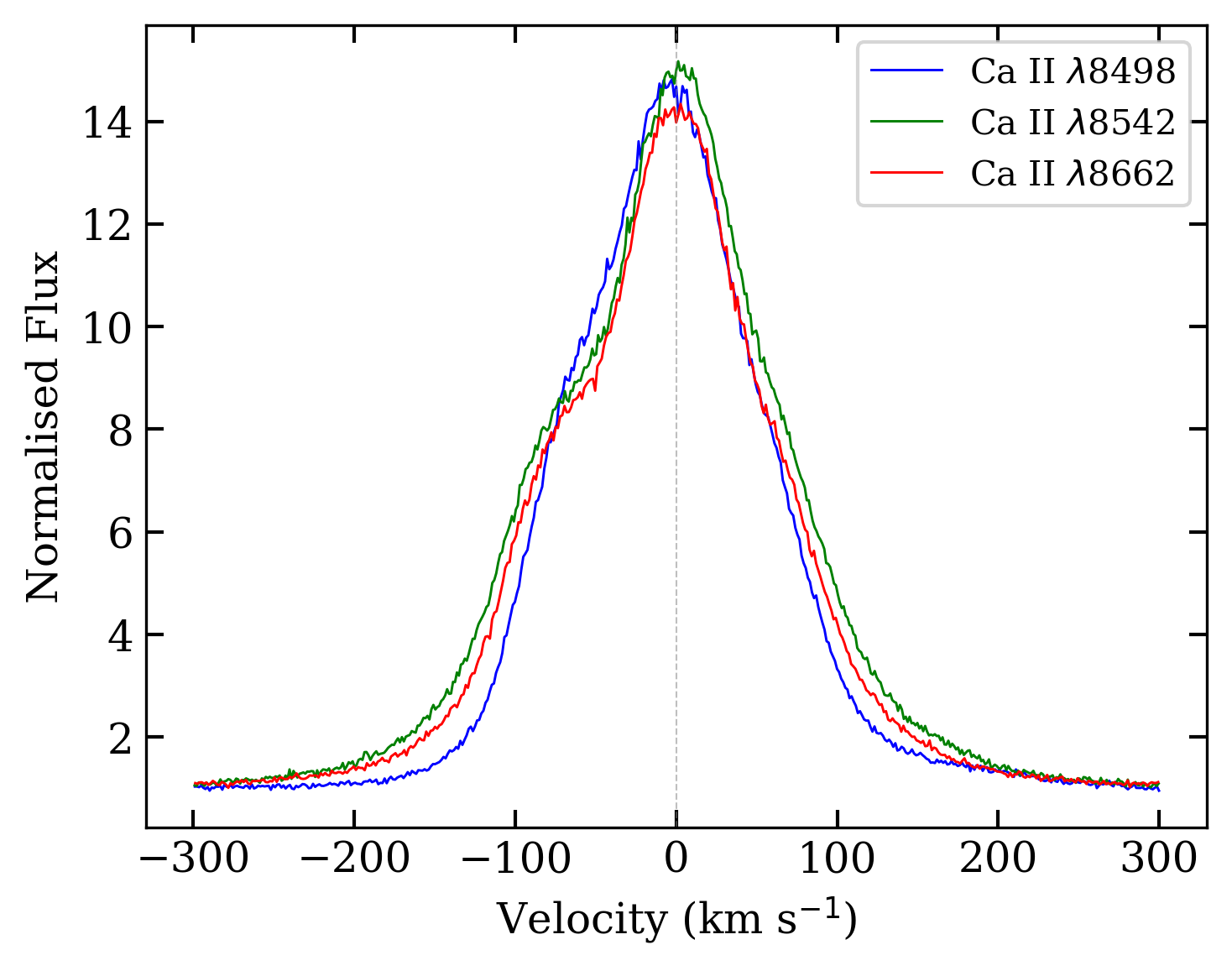}
    \includegraphics[width=0.49\linewidth]{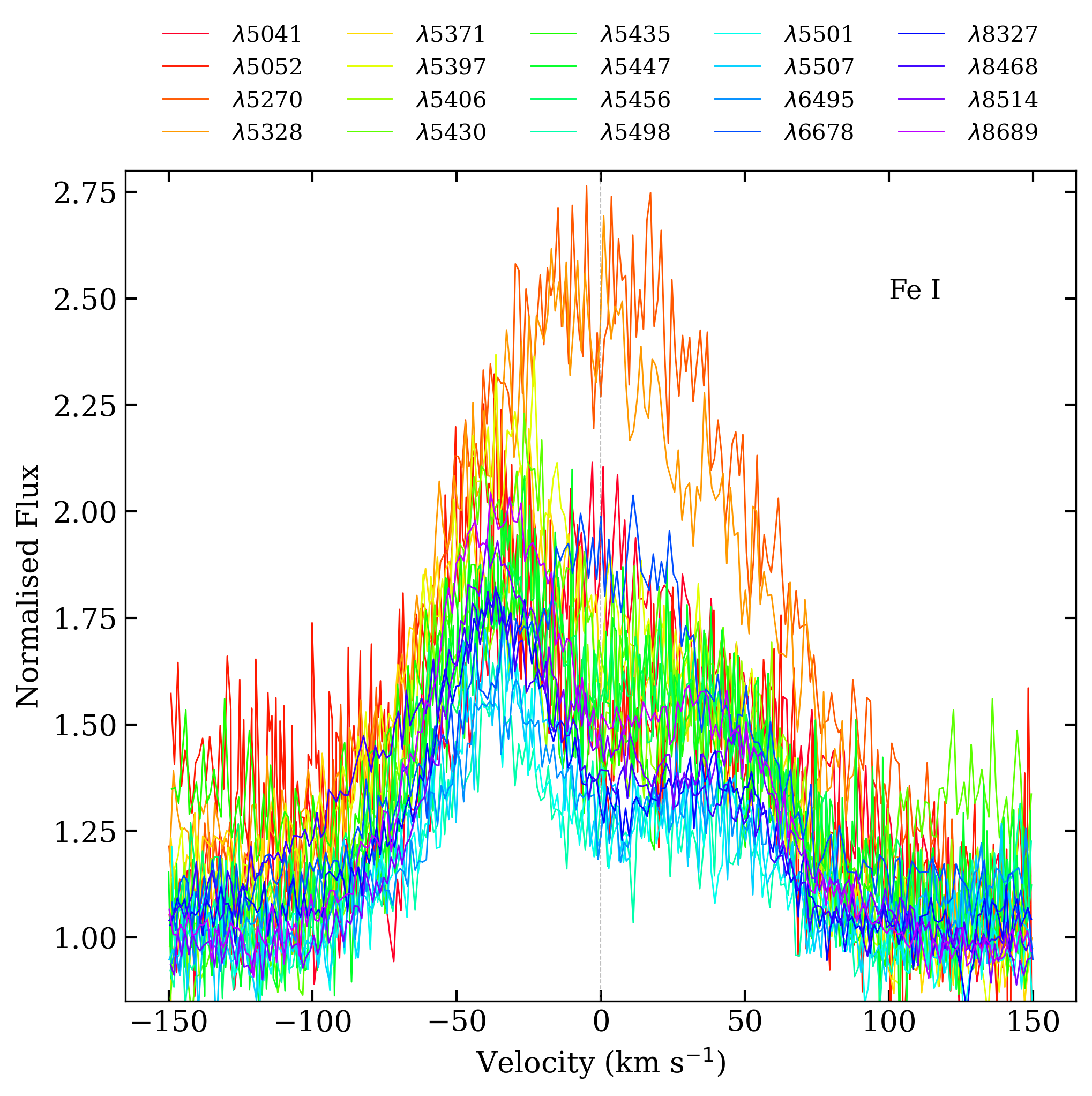}
    \includegraphics[width=0.49\linewidth]{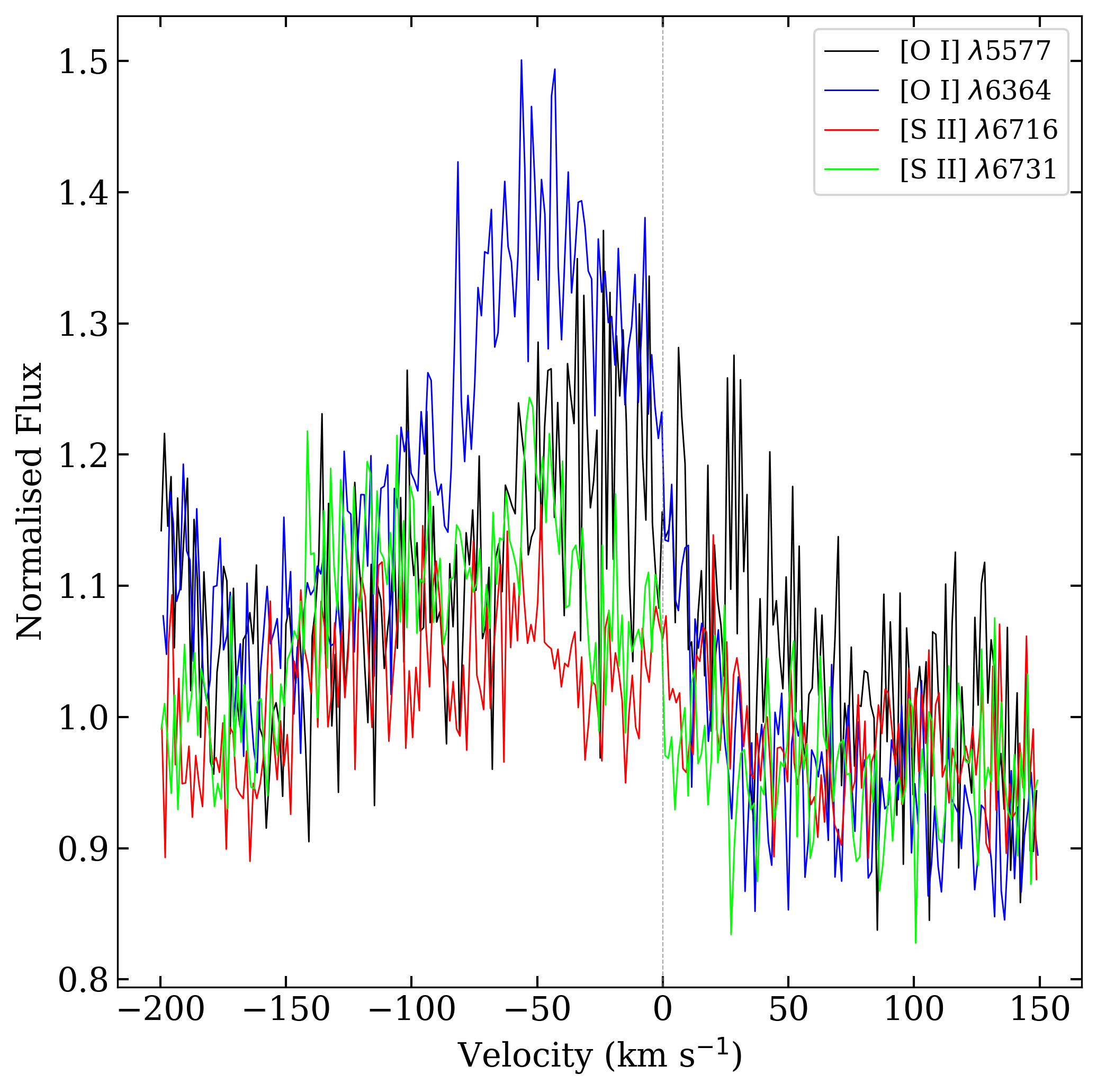}
    \includegraphics[width=0.49\linewidth]{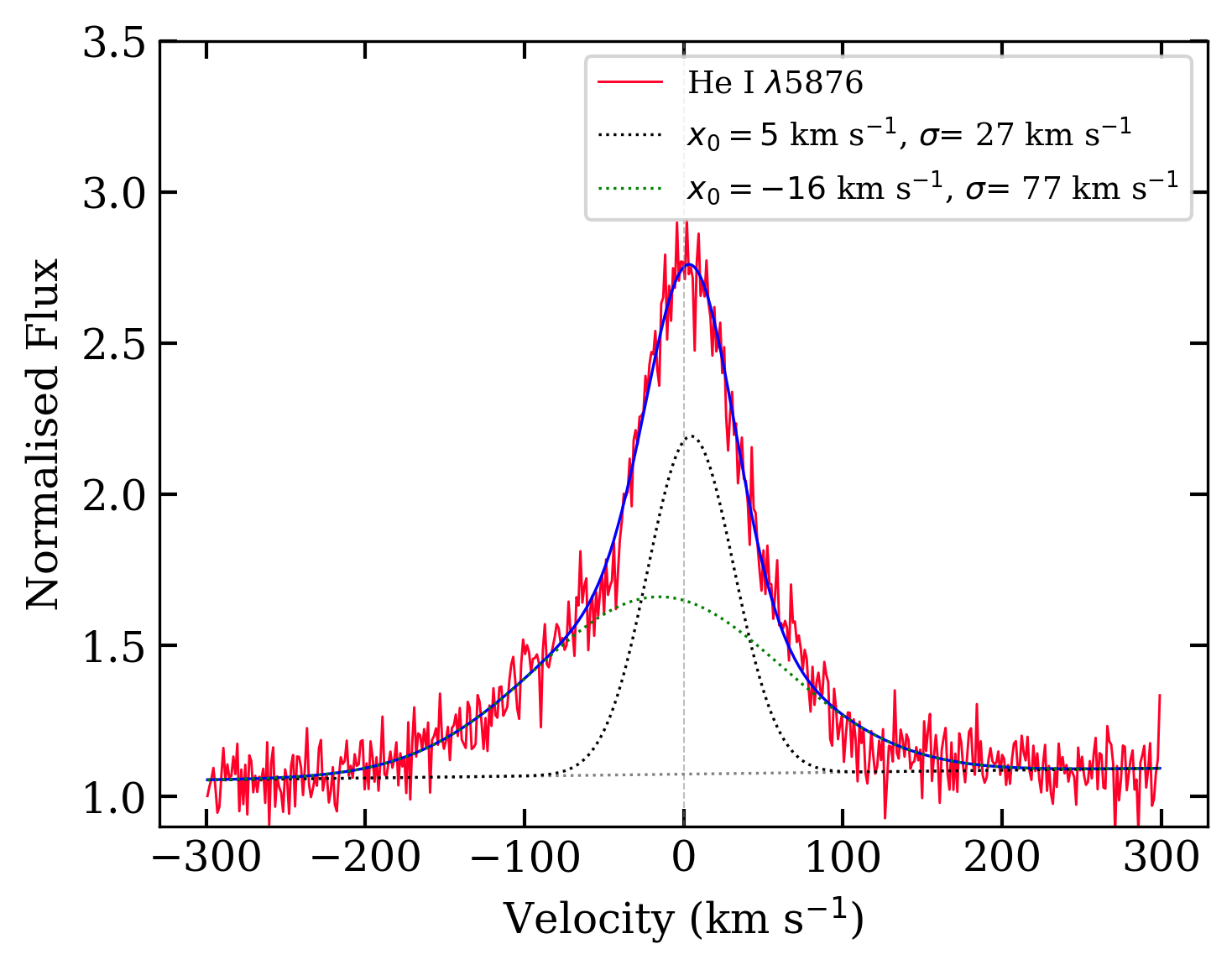}
    \centering
    \caption{Line profile of various detected lines in the HIRES spectrum of V1180 Cas. The He I $\lambda$5876 (\textit{lower panel}) profile is decomposed into two Gaussian components (black and green dotted curves), while the blue solid curve represents the best-fit composite profile.}
    \label{fig:line_profile}
\end{figure*}

The He I $\lambda$5876 emission profile in V1180 Cas is well described by two Gaussian components (see the lower panel of Figure \ref{fig:line_profile}). The NC is centered at $\sim$5 km s$^{-1}$ with FWHM $\sim$64 km s$^{-1}$. Given the spectral resolution ($\sim$4 km s$^{-1}$) and potential uncertainties in the wavelength calibration, this small velocity offset should be interpreted with caution. The BC is centered at $\sim$$-16$ km s$^{-1}$ with FWHM $\sim$180 km s$^{-1}$.

A two-component He I $\lambda$5876 profile of this kind is commonly observed in classical T Tauri stars \citep{Beristain_2001ApJ...551.1037B,Sicilia_2012A&A...544A..93S,Bouvier_2020A&A...643A..99B,2024A&A...690A.225A}.

\subsection{Radial Velocity} \label{subsec:rv}
The retrieved HIRES spectrum was already corrected for heliocentric motion during the standard reduction process. To determine the systemic velocity of V1180 Cas with respect to the Sun, we primarily used the Li I $\lambda$6708 absorption line \citep{Pascucci_2015ApJ...814...14P, Fang_2018ApJ...868...28F}, which originates in the stellar photosphere and serves as a well-established indicator of stellar youth \citep{Sestito_2008A&A...487..965S, 2021MNRAS.503..938Z}. The Li I $\lambda$6708 absorption line is the only clearly identified photospheric absorption feature in the HIRES spectrum. The non-detection of other photospheric absorption lines may be partly related to continuum veiling associated with accretion, which can weaken photospheric absorption features. We note, however, that the Li I $\lambda6708$ feature can be affected by nearby blends, stellar activity, veiling, and accretion-related effects, which may introduce systematic uncertainties in its centroid position \citep{Campbell_2023A&A...673A..80C}. As an independent consistency check, we additionally examined Fe I lines detected in emission in the HIRES spectrum. Fe I emission lines are associated with chromospheric activity in young stars and have been used as tracers of the stellar radial velocity (RV) \citep[e.g.,][]{Petrov_2011A&A...535A...6P}. However, chromospheric activity can introduce temporal modulation in the RV measured from these lines, similar to photospheric absorption lines \citep{Petrov_2011A&A...535A...6P}. From the numerous detected Fe I emission lines, we selected only those that are relatively isolated and exhibit symmetric profiles to ensure reliable velocity estimates.

\begin{figure*}
    \centering
    \includegraphics[width=0.49\linewidth]{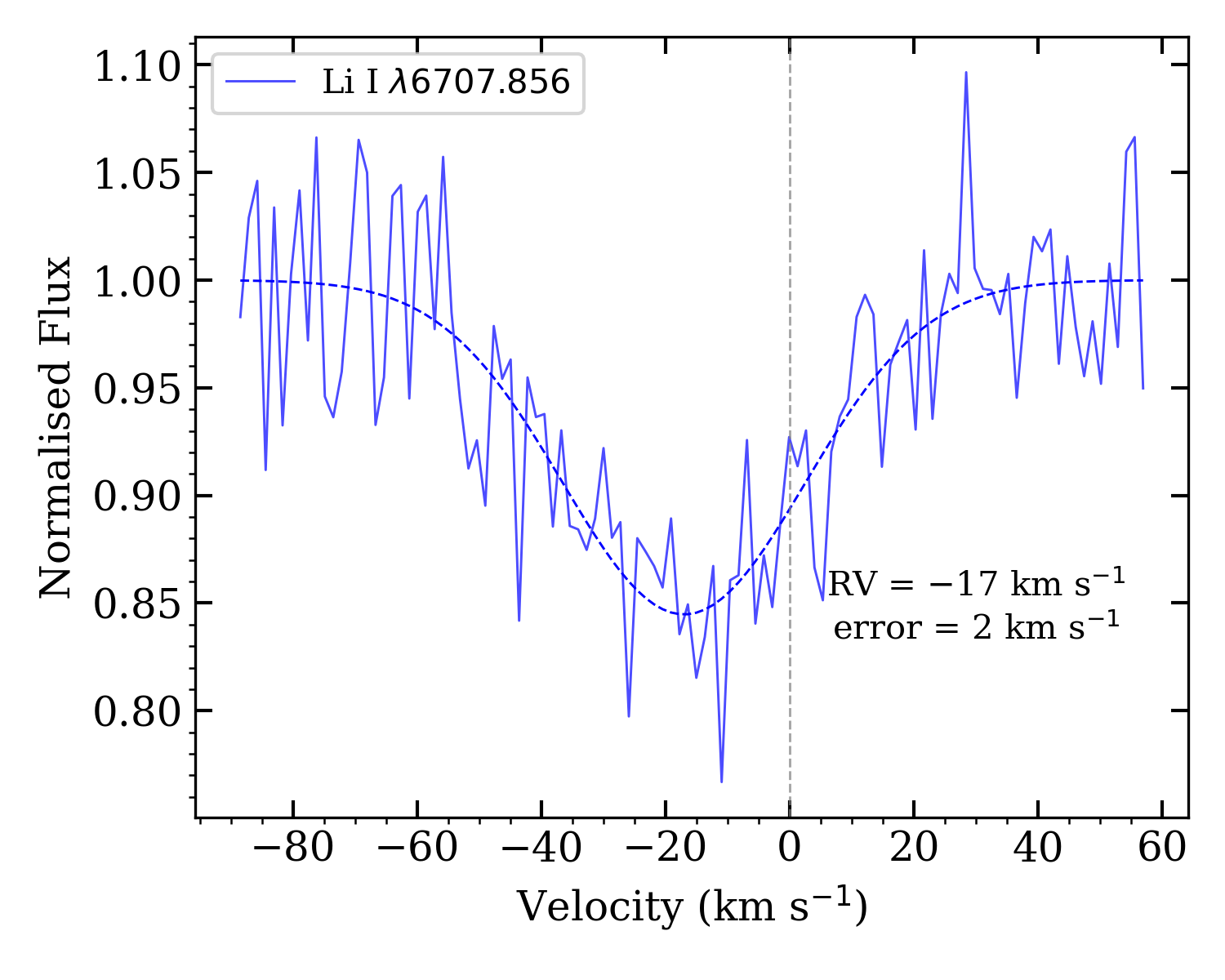}
    \includegraphics[width=0.49\linewidth]{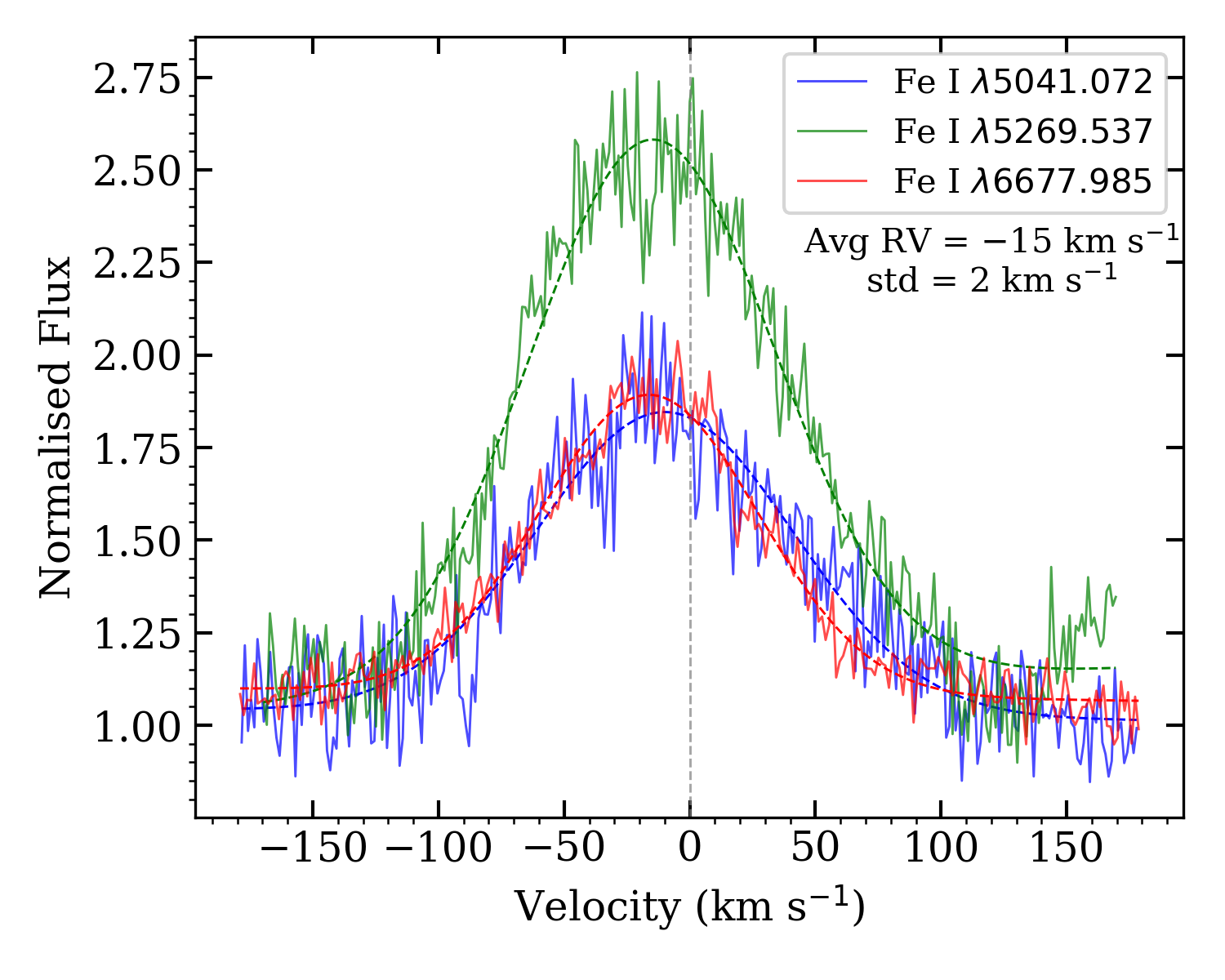}
    \centering
    \caption{RV measurement using Li I absorption line (left panel) and Fe I emission lines (right panel). The dashed lines indicate the best-fit Gaussian profile over observed data, and the vertical dashed grey line shows the expected line center.}
    \label{fig:rv}
\end{figure*}

Figure \ref{fig:rv} shows line profiles of the Li I line (left panel) and selected Fe I emission lines (right panel). Each profile was fitted with a Gaussian function, and the line centroid was used to derive the corresponding velocity. The uncertainty in each measured velocity was estimated from the uncertainty in the fitted Gaussian centroid. The wavelength-calibration rms of $\sim0.07$ \r{A} corresponds to $\sim3.1$ km s$^{-1}$ at the wavelength of the Li I line and is considered as an additional potential systematic uncertainty.

The Li I line yields a central velocity of $\sim -17 \pm 2$ km s$^{-1}$, while the average velocity derived from three Fe I lines is $\sim -15 \pm 2$ km s$^{-1}$. The average of these two measurements is adopted as the systemic RV of the V1180 Cas system, resulting in RV $= -16 \pm 3$ km s$^{-1}$, where the uncertainty is derived through propagation of the errors in individual measurements including systematic uncertainty associated with the wavelength calibration.

Converting to the Local Standard of Rest, this heliocentric RV corresponds to $V_{\rm LSR} \sim -13$ km s$^{-1}$. This value is broadly consistent with the systemic velocity of the parent molecular cloud, L1340, for which \citet{Kun_2003A&A...398..169K} report $V_{\rm LSR}$ in the range $-14.13$ to $-15.92$ km s$^{-1}$ from NH$_3$ core observations (their Table 1). The agreement (within $\sim$1$-$3 km s$^{-1}$) supports the physical association of V1180 Cas with the L1340 cloud and provides an additional, independent check on the reliability of our RV measurement. This is the first RV measurement reported for V1180 Cas.

\subsection{Line Profile of [O I] ($\lambda$6300)} 
\label{subsec:line_prof_OI}

\begin{figure}
    \centering
    \includegraphics[width=\linewidth]{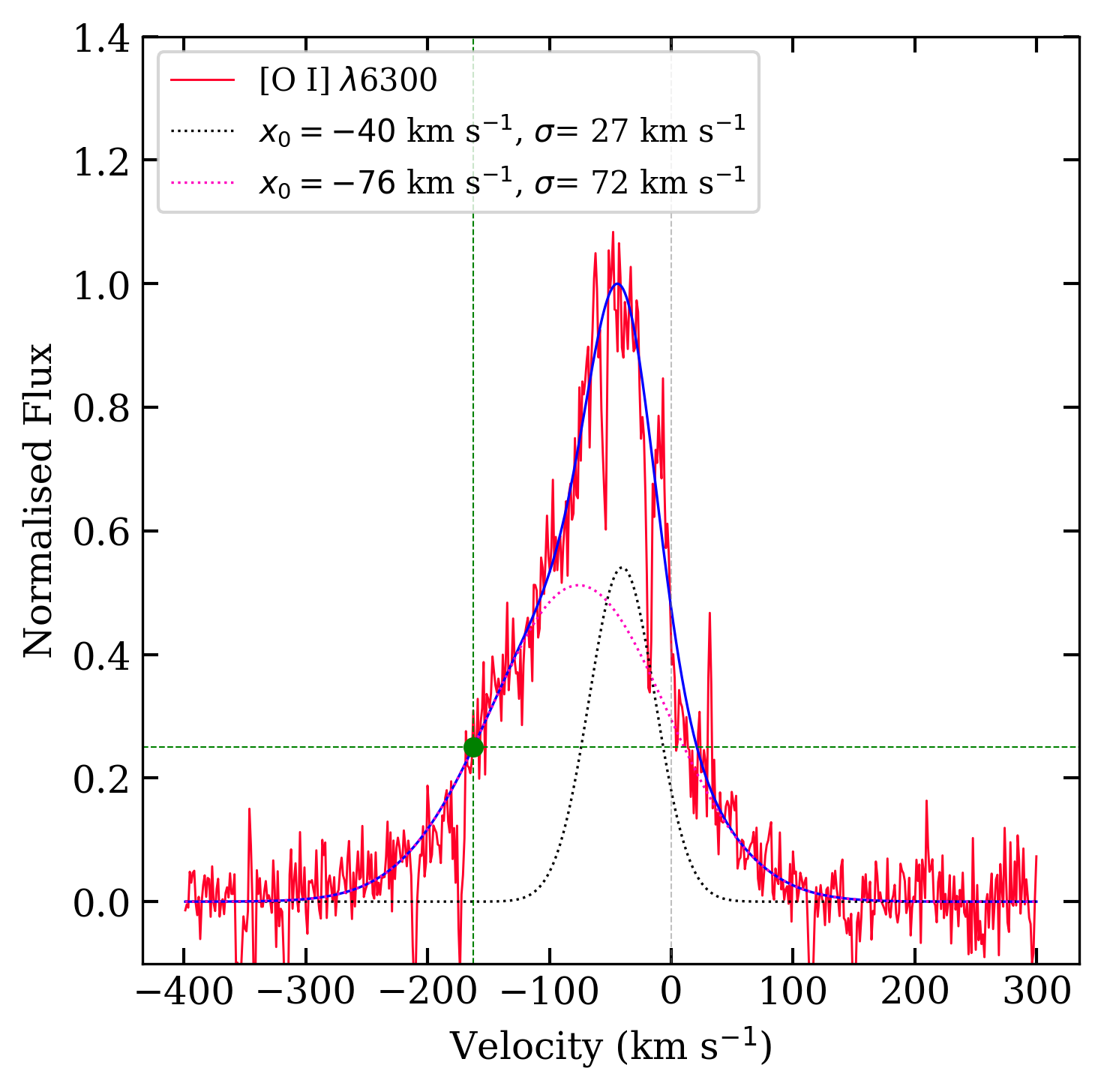}
    \centering
    \caption{Line profile of [O I] $\lambda$6300. The profile is continuum-subtracted and normalized to the maximum flux value. The black and magenta dotted curves represent the individual Gaussian components, centered at $\sim -40$ km s$^{-1}$ and $\sim -76$ km s$^{-1}$, respectively, while the blue solid curve shows the best-fit composite profile. The horizontal green dotted line marks 25\% of the maximum blue-wing flux, and the vertical green dotted line indicates the corresponding velocity of $-168$ km s$^{-1}$, which is used to estimate the inclination angle (see Section \ref{subsec:inc_ang}). The green dot shows the intersection point.} Regions affected by telluric absorption have been masked during the fitting procedure.
    \label{fig:oi}
\end{figure}

Figure \ref{fig:oi} shows the [O I] $\lambda$6300 line profile of V1180 Cas, which displays a clear asymmetry with an extended blue wing. Using multi-component Gaussian fitting, the line is well reproduced by two components: a HVC (shown in magenta dotted curve in Figure \ref{fig:oi}) with a centroid velocity of $\sim -76 \pm 2$ km s$^{-1}$ and FWHM $\sim 170 \pm5$ km s$^{-1}$, and a LVC (shown in black dotted curve in Figure \ref{fig:oi}) centered at $\sim -40 \pm1$ km s$^{-1}$ with FWHM $\sim 64\pm2$ km s$^{-1}$. The measured velocities are with respect to the stellar systemic velocity. Wavelength regions affected by telluric absorption were carefully identified and masked during the fitting procedure and were excluded from the determination of the line-profile parameters.

We note that the absolute velocities of the [O I] components are subjected to additional systematic uncertainties associated with the adopted systemic RV. In particular, \citet{Campbell_2023A&A...673A..80C} showed that the Li I $\lambda$6708 absorption feature in YSOs can exhibit source-to-source variations in its centroid position due to line blending, stellar activity, veiling, accretion-related velocity shifts, and star-spot coverage. They further demonstrated that using the Li I feature alone to establish the stellar RV may introduce systematic offsets of several km s$^{-1}$ in the velocities of emission lines such as [O I] $\lambda6300$.

The LVC centroid velocity measured for V1180 Cas is somewhat higher in magnitude than the values typically reported in recent high-resolution spectroscopic surveys of Class II YSOs \citep[e.g.,][]{Nisini_2018A&A...609A..87N, Fang_2018ApJ...868...28F, Banzatti_2019ApJ...870...76B}.

\subsection{Inclination Angle} \label{subsec:inc_ang}
We estimate the disk inclination of V1180 Cas using two empirical approaches based on the kinematics of forbidden-line emission. First, we employ the empirical relation presented by \cite{Appenzeller_2013A&A...558A..83A}, which connects the kinematics of forbidden-line emission to disk geometry. In that study, inclination angles for a sample of Class II YSOs were independently determined from spatially resolved millimeter and radio observations, and subsequently correlated with the kinematic properties of the [O I] $\lambda$6300 line, in particular the blue-velocity ($V_{\mathrm{be}}$) where intensity drops to 25\% of the peak value (shown by dotted green line in Figure \ref{fig:oi}).

\begin{figure*}
    \centering
    \includegraphics[width=0.49\linewidth]{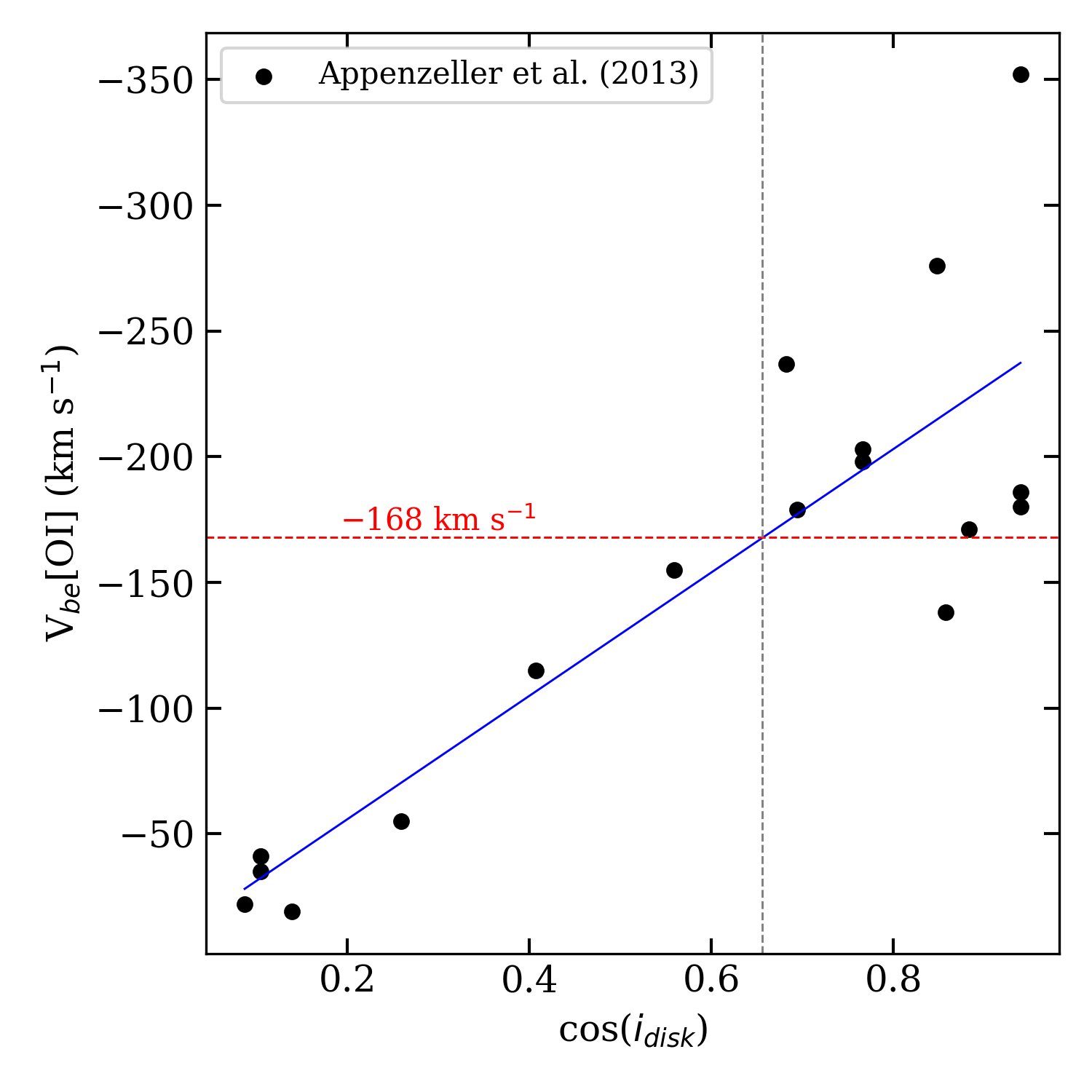}
    \includegraphics[width=0.49\linewidth]{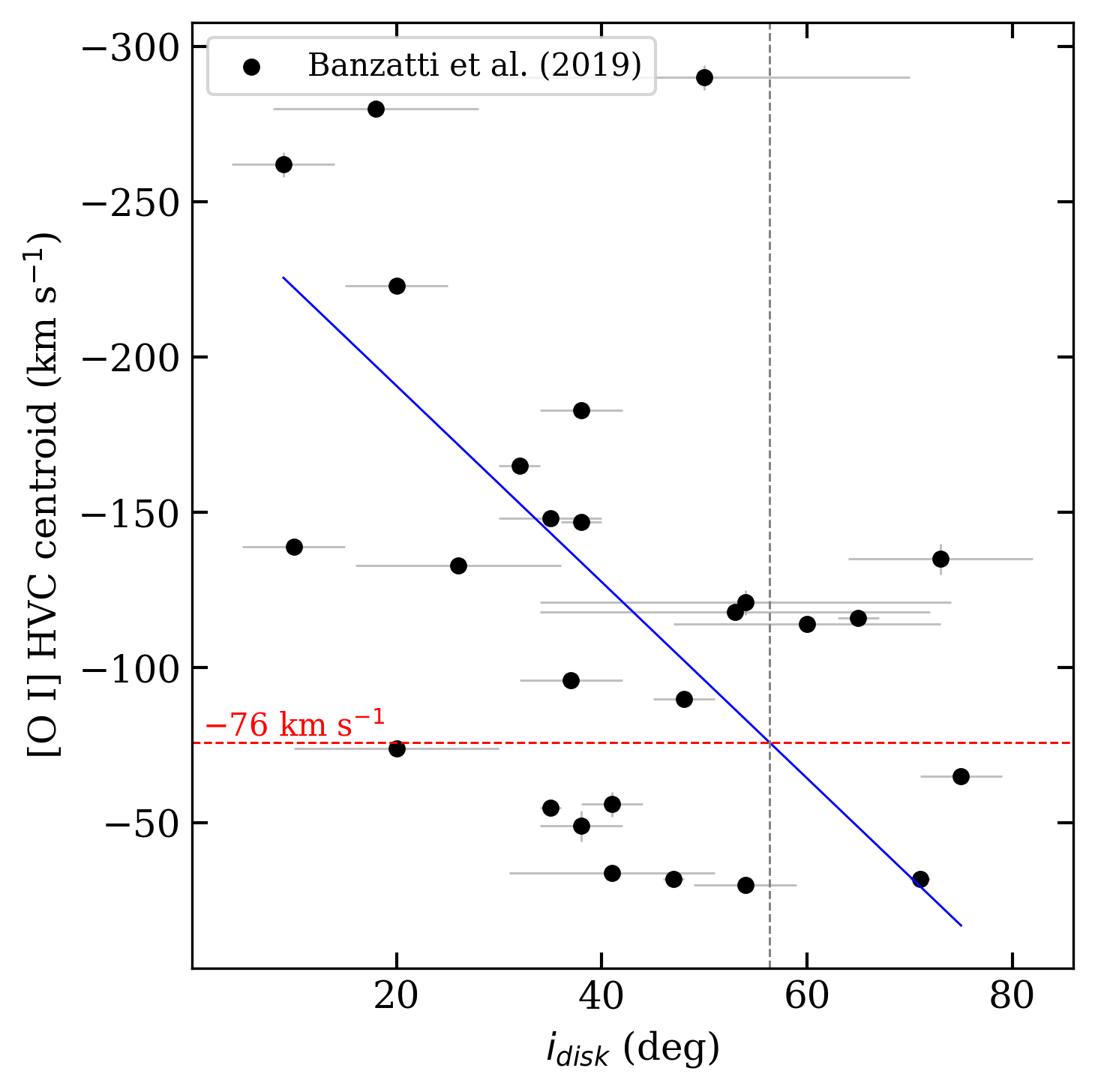}
    \centering
    \caption{The left panel illustrates the empirical relation between V$_{be}$ in [O I] $\lambda 6300$ (see text in Section \ref{subsec:inc_ang} for details) and $\cos(i_{disk})$ for Class II YSOs (data from Table 1 of \citealt{Appenzeller_2013A&A...558A..83A}); the horizontal red dashed line marks the measured V$_{be}$  for V1180 Cas and the vertical grey dashed line indicates the corresponding $\cos(i_{disk})$. The right panel shows the empirical relation between the centroid velocity of HVC in [O I] $\lambda 6300$ and $i_{disk}$ \citep{Banzatti_2019ApJ...870...76B}. The horizontal red dashed line indicates the measured HVC centroid velocity for V1180 Cas, and the vertical grey dashed line marks the corresponding $i_{disk}$. The blue solid lines in both panels represent linear fits to the respective data.}
    \label{fig:inc_ang}
\end{figure*}

We apply this method to the HIRES spectrum by measuring $V_{\mathrm{be}}$ from the [O I] $\lambda$6300 profile. The blue-edge velocity is found to be $V_{\mathrm{be}} \approx -168$ km s$^{-1}$ (see Figure \ref{fig:oi}). Using the empirical relation of \citet{Appenzeller_2013A&A...558A..83A}, this value corresponds to $\cos(i_{\mathrm{disk}}) = 0.66$, yielding a disk inclination of $i_{\mathrm{disk}} \approx 50^\circ$ (see left panel in Figure \ref{fig:inc_ang}). Considering the uncertainty associated with the empirical relation, we estimate an uncertainty of approximately $8^\circ$ on the derived inclination.

As an independent estimate, we also use the empirical correlation between the centroid velocity of the HVC of the [O I] $\lambda6300$ line and disk inclination reported by \citet{Banzatti_2019ApJ...870...76B}. The authors analyzed high-resolution optical spectra of 65 T Tauri stars, 32 of which exhibited an HVC, and reported a statistically significant correlation between HVC centroid velocity and disk inclination. For V1180 Cas, the centroid velocity of the [O I] $\lambda6300$ HVC is $-76\pm2$ km s$^{-1}$ (see Section \ref{subsec:line_prof_OI}). Applying the correlation of \citet{Banzatti_2019ApJ...870...76B} yields an inclination of $i_{\mathrm{disk}}=56\pm17^\circ$ (see right panel in Figure \ref{fig:inc_ang}). Here, the $\pm17^\circ$ uncertainty is obtained through standard error propagation, taking into account the uncertainty in the HVC centroid velocity and the uncertainties in the parameters of the linear fit reported by \citet{Banzatti_2019ApJ...870...76B}.
The two empirical approaches therefore provide comparable inclination estimates, $i_{\mathrm{disk}}\approx50\pm8^\circ$ and $56\pm17^\circ$, respectively. We emphasize that this estimate should be regarded as approximate because of the intrinsic scatter in the empirical correlations, the source-to-source variation in outflow velocities.
We also note that these relations are calibrated mainly using classical T Tauri stars, and their application to eruptive systems such as V1180 Cas may introduce additional systematic uncertainties.
 
UX Orionis (UXor) stars are young stellar objects that exhibit irregular, often pronounced photometric variability, commonly attributed to variable extinction by circumstellar dust; during deep fading events, the increasing contribution of scattered light can produce the characteristic "blueing" effect \cite{Herbst_1994AJ....108.1906H, Abraham_2018ApJ...853...28A}.
Radiative transfer models of UXor-type variability by \citet{Natta_2000A&A...364..633N} showed that the characteristic extinction-driven dimming and blueing effect can arise in systems viewed at intermediate-to-high inclination angles ($\sim45^{\circ}$--$68^{\circ}$), where circumstellar dust intermittently obscures the central star while scattered light becomes increasingly important during deep fading events. The inclination angle derived here ($\sim50^\circ$) falls within this range, and is consistent with the hybrid UXor/EXor nature proposed for V1180 Cas by \citet{Chand_2026ApJ...998..204C}.

\begin{table*}
\centering
\caption{Spectral line information. Line fluxes ($F_{line}$) listed are A$_{V}$ corrected.}
\begin{tabular}{ccccc}
\hline
\hline
        Line &Wavelength& EW &$F_{line}$ \\
             & (\r{A}) & (\r{A}) &($\times$10$^{-13}$ erg s$^{-1}$ cm$^{-2}$)\\
        \hline
        H$\beta$  & 4861.325 & $-36.04\pm0.25$ & $13.15\pm0.34$ \\
        $[$O I$]$     & 5577.340 & $-0.61\pm0.05$ & -- \\
        He I      & 5875.621 & $-4.39\pm0.09$ & $0.59\pm0.02$ \\
        $[$O I$]$     & 6300.304 & $-3.60\pm0.08$ & $0.78\pm0.02$ \\
        $[$O I$]$     & 6363.776 & $-0.79\pm0.05$ & -- \\
        H$\alpha$ & 6562.819 & $-194.12\pm1.21$ & $35.53\pm0.69$ \\
        $[$S II$]$    & 6716.440 & $-0.35\pm0.05$ & -- \\
        $[$S II$]$    & 6730.800  & $-0.55\pm0.06$ & -- \\
        O I       & 8446.360 & $-7.22\pm0.13$ & $0.70\pm0.02$ \\
        Ca II     & 8498.018 & $-54.47\pm0.48$ & $5.12\pm0.10$ \\
        Ca II     & 8542.089 & $-62.34\pm0.49$ & $5.73\pm0.11$ \\
        Ca II     & 8662.140 & $-57.21\pm0.50$ & $4.96\pm0.10$ \\
        
\hline
\end{tabular}
\label{tab:line_info}
\end{table*}

\begin{table*}
\centering
\caption{Line luminosities, accretion luminosities and mass accretion rates.}
\begin{tabular}{ccccc}
\hline
\hline
        Line &Wavelength&L$_{line}$   & L$_{acc}$   &  $\dot{M}_{acc}$ \\
             & (\r{A})  &(L$_{\odot}$) &(L$_{\odot}$)& ($\times$10$^{-7}$ $M_{\odot}$ yr$^{-1}$)\\
        \hline
        H$\beta$  & 4861.325 & $(3.3\pm0.1)\times10^{-2}$ & $7.7 \pm 0.2$ & $23.6\pm0.8$ \\
        He I      & 5875.621 & $(1.0\pm0.1)\times10^{-3}$ & $2.5 \pm 0.1$ & $7.7\pm0.3$ \\
        H$\alpha$ & 6562.819 & $(8.7\pm0.2)\times10^{-2}$ & $3.5 \pm 0.1$ & $10.6\pm0.2$ \\
        O I       & 8446.360 & $(2.0\pm0.1)\times10^{-3}$ & $2.9 \pm 0.1$ & $9.0\pm0.3$ \\
        Ca II     & 8498.018 & $(1.2\pm0.1)\times10^{-2}$ & $5.2 \pm 0.1$ & $15.9\pm0.3$ \\
        Ca II     & 8542.089 & $(1.4\pm0.1)\times10^{-2}$ & $4.3 \pm 0.1$ & $13.1\pm0.2$ \\
        Ca II     & 8662.140 & $(1.2\pm0.1)\times10^{-2}$ & $3.3 \pm 0.1$ & $10.1\pm0.3$ \\
        
\hline
\end{tabular}
\label{tab:ac_rates}
\end{table*}

\subsection{Equivalent Widths and Line Fluxes}
\label{subsec:EW_LF}
The equivalent widths (EWs) and line fluxes are measured for several spectral lines identified in the HIRES spectrum. For each line, spectral cutouts are extracted by manually selecting windows centered on the line profile, ensuring that the full extent of the emission feature is included: $\pm$15 \r{A} for H$\alpha$ and $\pm$10 \r{A} for the other lines. The local continuum was determined from a linear fit to data points within 3 \r{A} of both edges of each window. These wavelength intervals are selected through visual inspection to avoid contamination from the emission-line wings and nearby spectral features. After continuum subtraction, the residual line profile is integrated using the composite trapezoidal rule to obtain the EW. 

Since the HIRES spectrum is not flux calibrated, continuum flux densities are required to convert the measured EWs into line fluxes. We therefore use available photometric measurements in the $V$ ($\lambda_{\rm eff}\approx5450$ \r{A}), $R$ ($\lambda_{\rm eff}\approx6410$ \r{A}), and $I$ ($\lambda_{\rm eff}\approx7980$ \r{A}) bands, whose effective wavelengths are close to those of the respective emission lines. The continuum flux densities were derived from $VRI$ photometric observations obtained on 2016 August 06 by \citet{Mutafov_2022RAA....22l5014M}, when V1180 Cas had $V=17.85\pm0.03$, $R=16.45\pm0.02$, and $I=15.44\pm0.02$ mag. These observations were obtained 12 days before the HIRES spectrum on 2016 August 18. The next available $VRI$ photometric observations were obtained on 2016 October 02, when the source had $V=17.84\pm0.03$, $R=16.36\pm0.02$, and $I=15.33\pm0.02$ mag \citep{Mutafov_2022RAA....22l5014M}. The similar brightness at these two epochs, together with the long-term light curve showing that V1180 Cas remained in a relatively stable bright state from January 2016 to January 2017, with a variability amplitude of $\Delta V<0.5$ mag \citep{Mutafov_2022RAA....22l5014M, Chand_2026ApJ...998..204C}, suggests that the source brightness did not change significantly around the HIRES observing epoch. We therefore use the 2016 August 06 photometry to estimate the continuum flux densities at the spectroscopic epoch. The EWs were converted to line fluxes by multiplying them by the corresponding continuum flux densities.

V1180 Cas exhibits variable extinction; therefore, the choice of extinction appropriate to the spectroscopic epoch is important. The HIRES spectrum analyzed here was obtained on 2016 August 18, when the source was in a bright and relatively stable photometric state. However, the HIRES spectrum is not flux calibrated and shows only one clearly identifiable photospheric absorption feature, Li I $\lambda$6708, while other expected photospheric absorption features are not clearly detected. This may be due to accretion-related veiling, which can weaken or mask relatively weak photospheric absorption lines. Consequently, the HIRES spectrum does not provide a reliable independent constraint on $A_V$. We therefore retain $A_V=4.6$ mag, adopted by \citet{Chand_2026ApJ...998..204C} from the SED-based analysis of \citet{Kun_2016ApJ...822...79K}, which used SDSS, 2MASS, and Spitzer photometry. This value is also broadly consistent with the independent estimate of $A_V=4.3$ mag obtained by \citet{Kun_2011} from the observed $I_C-J$ color, assuming a K7 spectral type. The line fluxes were corrected for extinction using the extinction law of \citet{Cardelli_1989ApJ...345..245C} with $R_V=3.1$. The derived EWs and line fluxes are listed in Table \ref{tab:line_info}.

Since V1180 Cas exhibits variable extinction and the extinction appropriate to the HIRES epoch cannot be independently constrained from the available spectrum, the uncertainty in the adopted extinction introduces a systematic uncertainty in the derived line luminosities. This uncertainty propagates directly into both the accretion and outflow estimates, as the extinction-corrected line luminosities are used to derive these quantities (see Section \ref{subsec:outflow}). The effect is wavelength-dependent and is more pronounced for the shorter-wavelength optical lines.

\subsection{Distance to V1180 Cas}
\label{subsec:distance}

\citet{Bailer_2021AJ....161..147B} derived a Gaia DR3 distance of $883^{+44}_{-38}$ pc for V1180 Cas. The Gaia DR3 astrometric solution has a renormalized unit weight error (RUWE) of 1.066, which is below the threshold of 1.22 suggested for single objects by \citet{Castro_2024A&A...688A...1C}, indicating a good-quality astrometric solution and supporting the reliability of the Gaia parallax. The Gaia-based distance is also consistent with the revised distance range of 750--940 pc estimated for the L1340 cloud by \citet{Kun_2016ApJ...822...79K}. We therefore adopt $d=883$ pc for V1180 Cas throughout this work, including the calculations of accretion and outflow rates.

V1180 Cas is located in the Lynds 1340 (L1340) molecular cloud $(l,b)=(130^{\circ}.1,11^{\circ}.5)$. The distance to L1340 was initially estimated to be $\sim600$ pc by \citet{Kun_1994A&A...292..249K}. This estimate was subsequently revised by \citet{Kun_2016ApJ...822...79K}. The authors used spectroscopic and photometric observations of two of the most luminous stars associated with the cloud and compared their properties with intermediate-mass pre-main-sequence evolutionary models of \citet{Palla_1993ApJ...418..414P}. They obtained a distance range of 750--940 pc, considering a $\pm1$ subclass uncertainty in the spectral types, and adopted a distance of 825 pc for the L1340 cloud.

\subsection{Accretion Rate}
\label{subsec:accrntn}
We estimate the mass accretion rate of V1180 Cas using several permitted emission lines, including H$\beta$, H$\alpha$, He I $\lambda5876$, O I $\lambda8446$, and the Ca II IRT emission lines. The line luminosities are calculated from the extinction-corrected line fluxes using a distance of $d=883$ pc \citep{Bailer_2021AJ....161..147B}. The accretion luminosities ($L_{acc}$) are then derived from the line luminosities using the empirical relation from \cite{Alcala_2014A&A...561A...2A, Alcala_2017A&A...600A..20A}. Finally, the mass accretion rate is calculated using the relation:
\begin{equation} 
\label{eq:accrntn_rate}
\dot{M}_{\text{acc}} = \frac{L_{\text{acc}} R_{\ast}}{G M_{\ast} \left(1 - \frac{R_{\ast}}{R_{\text{in}}}\right)} 
\end{equation}

The stellar parameters adopted in this calculation are based on the characterization of V1180 Cas by \citet{Kun_2016ApJ...822...79K}. Using spectral energy distribution (SED) fitting and Hertzsprung--Russell diagram, \citet{Kun_2016ApJ...822...79K} derived a spectral type of approximately K7, with an effective temperature of T$_{eff}=3970^{+170}_{-90}$ K, a stellar luminosity of L$_{\ast}=3.19^{+2.25}_{-1.32}$ $L_{\odot}$, and the stellar mass of M$_{\ast}=0.65$ $M_{\odot}$. Using the derived T$_{eff}$ and L$_{\ast}$, we estimate a stellar radius of R$_{\ast}\approx5$ $R_{\odot}$ and we adopt an inner disk truncation radius of R$_{in}=5$ R$_{\ast}$. Here in Equation \ref{eq:accrntn_rate}, $G$ is the gravitational constant.

The derived line luminosities and corresponding mass accretion rates obtained using Equation \ref{eq:accrntn_rate} from H$\beta$, H$\alpha$, He I $\lambda5876$, O I $\lambda8446$, and the Ca II IRT lines are listed in Table \ref{tab:ac_rates}.

The derived accretion rates are higher than those reported in previous studies for V1180 Cas during similarly bright states ($I\approx15$ mag), including $\sim1.6\times10^{-7}$ $M_{\odot}$ yr$^{-1}$ from the Ca II $\lambda8542$ line by \citet{Kun_2011}, $1.58\times10^{-6}$ $M_{\odot}$ yr$^{-1}$ from H$\alpha$ by \citet{Kun_2016ApJ...822...79K}, and $(2.6$--$3.8)\times10^{-7}$ $M_{\odot}$ yr$^{-1}$ from H$\alpha$ and Ca II IRT by \citet{Chand_2026ApJ...998..204C}. In the present analysis, the accretion rates span $(7.7$--$23.6)\times10^{-7}$ $M_{\odot}$ yr$^{-1}$, with the spread reflecting the different line tracers and their empirical calibrations. The overall higher accretion rates obtained in this study are primarily attributable to the revised stellar parameters, particularly the adopted distance and stellar radius. Nevertheless, the derived rates remain within the range reported for strongly accreting EXor-type YSOs \citep{Fischer_2023ASPC..534..355F, Miera_2023A&A...678A..88C, Singh_2024ApJ...968...88S}.

\subsection{Density, Temperature, and Mass Outflow Rate}
\label{subsec:outflow}
The [S II] $\lambda6717/\lambda6731$ ratio is 0.64, which corresponds to an electron density of $\sim4 \times 10^{3}$ cm$^{-3}$ \citep{Osterbrock_2006agna.book.....O}. The [O I] $\lambda5577/\lambda6300$ ratio of 0.17 implies an electron density of $\sim4 \times 10^{6}$ cm$^{-3}$ for a gas temperature of $10^{4}$ K \citep{Natta_2014A&A...569A...5N}. Adopting this density, the ratio [O I] $\lambda5577/(\lambda6300+\lambda6364)\sim0.15$ yields a characteristic temperature of $\sim10^{4}$ K \citep{Osterbrock_2006agna.book.....O}. These density and temperature estimates were verified using the PyNeb Python package \citep{pyneb_2015A&A...573A..42L}.

The line luminosities of the HVC and LVC components of the [O I] $\lambda6300$ line are derived from the two-component Gaussian fits described in Section \ref{subsec:line_prof_OI}. The component line fluxes are corrected for extinction as described in Section \ref{subsec:EW_LF} and the corresponding line luminosities are calculated adopting a distance of $d=883$ pc \citep{Bailer_2021AJ....161..147B}. For the HVC, we estimate the jet mass-loss rate using Equation (12) of \citet{Fang_2018ApJ...868...28F}. We adopt a jet velocity in the sky plane $V_{\perp}=100$ km s$^{-1}$, an electron density $n_e=5\times10^{4}$ cm$^{-3}$, and a gas temperature of $T=10^4$ K following \citet{Nisini_2018A&A...609A..87N, Fang_2018ApJ...868...28F}.  
In this prescription, the projected size of the spectroscopic aperture is used as the characteristic transverse scale $l_{\perp}$. For the 0.574$^{\prime\prime}$ HIRES slit and the adopted distance of 883 pc, this corresponds to a projected aperture size of $\sim507$ au. We emphasize that this slit width should not be interpreted as the physical width of the jet. The actual transverse extent and filling factor of the HVC within the slit are not constrained by our observations. Thus, the aperture-based assumption introduces an additional systematic uncertainty in the derived mass-loss rate, particularly because the jet may not fully occupy the spectroscopic aperture. Using the extinction-corrected HVC luminosity of the [O I] $\lambda6300$ line, we obtain a jet mass-loss rate of $\dot{M}_{jet}$$\sim2.6 \times 10^{-8}$ $M_{\odot}$ yr$^{-1}$. Given the uncertainty in the jet filling factor and aperture geometry, this value should be regarded as an order-of-magnitude estimate.

For the LVC, we estimate the wind mass-loss rate using Equation (12) of \citet{Fang_2018ApJ...868...28F}. We adopt a disk inclination of $i_{disk}=50^{\circ}$, a projected wind velocity of $V_{wind}=40$ km s$^{-1}$, and a wind-emitting vertical extent of $l_{wind}=3.3$ au. The latter is estimated from the LVC FWHM of 64 km s$^{-1}$ by assuming a scaling factor of $f=10$ between the characteristic wind-launching radius and the vertical extent of the emitting region \citep{Fang_2018ApJ...868...28F}. Assuming $T=10^4$ K, we derive a wind mass-loss rate of $\dot{M}_{wind}\sim1.8\times10^{-8}$ $M_{\odot}$ yr$^{-1}$.

A comparison of these mass-loss rates with the mass accretion rate derived in Section \ref{subsec:accrntn}, and with values reported for other Class II YSOs, is presented in Section \ref{sec:disc}.
 
%%%%%%%%%%%%%%%%%%%%%%%%%%%%%%%%%%%%%%%%%%%%%%%%%%%%%%%%%%%%%%%%%%%%%%%%%%%%%%
\section{Interpretation and Discussion}
\label{sec:disc}

\subsection{Origin of the Permitted Line Emission}
The H$\alpha$ and H$\beta$ lines are known to form in hot accretion spots on the stellar surface and within the magnetospheric accretion columns, with possible additional contributions from chromospheric activity and a disk wind launched from the inner disk region \citep{Natta_2004A&A...424..603N, Lima_2010A&A...522A.104L, Kurosawa_2012MNRAS.426.2901K, Fang_2013ApJS..207....5F, Alcala_2014A&A...561A...2A, Ninan_2015ApJ...815....4N}. The blueshifted absorption components seen in both lines point to outflowing material along the line of sight.

The Ca II IRT emission in YSOs is commonly associated with accretion-related processes and has been widely used as an accretion diagnostic \citep{Muzerolle_1998AJ....116..455M, 2005ApJ...626..498M, Alcala_2014A&A...561A...2A, Alcala_2017A&A...600A..20A, Micolta_2024ApJ...976..251M}. The Ca II IRT intensity ratios measured for V1180 Cas suggest that these lines are formed predominantly in optically thick emitting regions \citep{Herbig_1980ApJ...242..628H, Hamann_1992ApJS...82..247H, Azevedo_2006, Alcala_2014A&A...561A...2A}.

Permitted Fe I emission lines are frequently observed in actively accreting YSOs and EXor-type \citep{Herbig_1977ApJ...217..693H, Herbig_2008AJ....135..637H, Ninan_2015ApJ...815....4N} variables, where they are believed to originate in dense, warm gas located in the inner circumstellar environment, particularly in the accretion-heated inner disk atmosphere \citep{Hamann_1992ApJS...82..247H, Sicilia_2012A&A...544A..93S, 2011ApJ...736...72K}. Most of the detected Fe I lines exhibit asymmetric profiles, while a few appear relatively symmetric. One possible explanation for this asymmetry is self-absorption, which would introduce a central dip in an otherwise symmetric profile, with symmetric lines expected to be comparatively optically thin. However, the relatively symmetric line at 5269.537 \r{A} in fact exhibits the highest relative optical depth of the sample, while several lower-$\tau$ lines (e.g., $\lambda$8468, $\lambda$8514) display pronounced asymmetry. This indicates that optical depth effects and self-absorption alone cannot account for the observed asymmetry in the Fe I profiles, suggesting that additional kinematic effects may contribute. If the double-peaked profiles were produced solely by rotational motion in a symmetric emitting region, one would generally expect the two peaks to be approximately symmetric and to have comparable intensities. However, the Fe I profiles in V1180 Cas often show unequal peak strengths. Therefore, the observed double-peaked profiles cannot be straightforwardly attributed to simple rotational broadening. In particular, the excess emission observed in the blue peak of several Fe I profiles may be associated with an additional blueshifted component, potentially arising from a hot disk wind. However, this interpretation remains tentative, as the present data do not allow us to uniquely distinguish between wind emission and other kinematic components within the inner accretion environment.

The He I $\lambda$5876 profile can be naturally interpreted in terms of two physically distinct components. The NC, close to the systemic velocity and having a FWHM of $\sim64$ km s$^{-1}$, is generally attributed to emission from the post-shock region at the base of the accretion columns, where material guided by magnetic field lines impacts the stellar surface \citep{Beristain_2001ApJ...551.1037B, Nowacki_2023A&A...678A..86N}. The narrow width and proximity to the systemic velocity of the NC are consistent with gas that has been decelerated after the accretion shock and confined to a compact region \citep{Beristain_2001ApJ...551.1037B}. The BC traces a more dynamic and spatially extended region: its larger width ($\sim180$ km s$^{-1}$) indicates a wide range of velocities, while the blueshifted centroid suggests a net outflowing motion along the line of sight. Such broad, slightly blueshifted He I emission profiles have been associated with both magnetospheric accretion flows and outflowing gas in the inner star--disk environment \citep{Beristain_2001ApJ...551.1037B, Kwan_2011MNRAS.411.2383K, Nowacki_2023A&A...678A..86N}; however, the present line profile alone does not allow us to uniquely distinguish between these contributions.

\subsection{A Coupled Accretion--Outflow System}
Taken together, the strong hydrogen lines (e.g.,  H$\beta$, H$\alpha$, and Paschen lines), Ca II IRT emissions indicate ongoing mass transfer from the inner disk onto the stellar surface via dense, ionized accretion columns, while the detection of multiple forbidden transitions, particularly [O I] $\lambda$6300, provides clear evidence of outflow activity \citep{Muzerolle_1998AJ....116..455M, White_2003ApJ...582.1109W, Alcala_2014A&A...561A...2A, Ninan_2015ApJ...815....4N}. The simultaneous detection of these diagnostics suggests that both accretion and mass-loss processes are active in V1180 Cas and is consistent with the possibility that they are physically linked, as proposed in magnetospheric accretion models for YSOs \citep{Bouvier_2007prpl.conf..479B, Hartmann_2016ARA&A..54..135H}. The disk inclination derived in Section \ref{subsec:inc_ang} ($i_{\mathrm{disk}} \approx 50^\circ$) suggests that the system is viewed at an intermediate inclination, consistent with the simultaneous presence of both accretion and outflow signatures in the observed line profiles.

The two-component He I profile may further support this interpretation, with the NC and BC likely tracing the accretion shock and the inner disk wind/magnetospheric flow, respectively.

Combining the derived jet mass-loss rate ($\dot{M}_{jet}\sim2.6\times10^{-8}$ $M_{\odot}$ yr$^{-1}$; Section \ref{subsec:outflow}) with the mass accretion rates derived in Section \ref{subsec:accrntn} (Table \ref{tab:ac_rates}), we obtain $\dot{M}_{jet}/\dot{M}_{acc}\sim0.01$--$0.03$ for V1180 Cas, i.e., the jet carries away only $\sim$1--3\% of the accreted mass. Including the wind mass-loss rate derived from the LVC ($\dot{M}_{wind}\sim1.8\times10^{-8}$ $M_{\odot}$ yr$^{-1}$), the combined jet+wind mass-loss corresponds to $\sim$2--6\% of $\dot{M}_{acc}$ -- roughly two orders of magnitude smaller than the accretion rate. This ratio lies toward the lower end of the broad range reported for Class II YSOs, $\dot{M}_{jet}/\dot{M}_{acc}\sim0.01$--$0.7$ \citep{Hartigan_1995ApJ...452..736H, Fang_2013ApJS..207....5F, Nisini_2018A&A...609A..87N}, though it remains consistent with expectations from magneto-centrifugal jet-launching models, in which only a fraction of the accreting material is redirected into collimated outflows while the bulk continues onto the star. These mass-loss estimates are also broadly consistent with previous, lower-resolution measurements of V1180 Cas \citep{Antoniucci_2014A&A...565L...7A, Chand_2026ApJ...998..204C}. Using the [O I] $\lambda6300$ line, \citet{Antoniucci_2014A&A...565L...7A} estimated a mass-loss rate of $\sim3\times10^{-9}$ $M_{\odot}$ yr$^{-1}$, while \citet{Chand_2026ApJ...998..204C} obtained a median mass-loss rate of $\sim1.2\times10^{-8}$ $M_{\odot}$ yr$^{-1}$ from observations obtained at different epochs between 2015 and 2022. The lower value reported by \citet{Antoniucci_2014A&A...565L...7A} may partly reflect the different observational state of the source: their observation on 2014 February 19 was obtained when V1180 Cas was $\sim0.6$ mag fainter in the $R$ band than during the HIRES epoch and was recovering from the deep photometric dip D4 (see Figure 2 of \citealt{Chand_2026ApJ...998..204C}). The higher mass-loss rate derived from the present high-resolution spectrum may partly reflect temporal variations in the outflow activity of V1180 Cas. However, the large systematic uncertainties associated with the adopted emitting-region size and filling factor may also contribute to the higher estimated value.

\subsection{Kinematics and Physical Conditions in the Outflowing Gas}
The clear separation of the [O I] $\lambda$6300 profile into LVC and HVC components indicates a stratified outflow. The relatively narrow and weakly blueshifted LVC is consistent with a slow disk wind, while the broader and more strongly blueshifted HVC is consistent with a faster, collimated jet \citep{Bacciotti_1999A&A...350..917B, Ferreira_2006A&A...453..785F, Agra-Amboage_2011A&A...532A..59A, Fang_2018ApJ...868...28F}. The relatively high LVC centroid velocity compared with typical T Tauri stars may indicate an origin closer to the central star, such as an inner-disk or stellar wind \citep{Banzatti_2019ApJ...870...76B}.

The forbidden-line ratios further indicate a density-stratified outflow. The [S II] ratio traces lower-density gas with $n_e\sim4\times10^{3}$ cm$^{-3}$, whereas the [O I] ratios probe denser gas with $n_e\sim4\times10^{6}$ cm$^{-3}$ and $T\sim10^{4}$ K. These values are broadly consistent with previous estimates for V1180 Cas \citep{Antoniucci_2014A&A...565L...7A, Chand_2026ApJ...998..204C} and with the range reported for other T Tauri outflows. For example, shock models and forbidden-line diagnostics typically yield densities of $\sim10^{3}$--$10^{4}$ cm$^{-3}$ from [S II] and $\sim10^{6}$--$10^{8}$ cm$^{-3}$ from [O I] diagnostics \citep{Hartigan_1995ApJ...452..736H, Natta_2014A&A...569A...5N, Oh_2015JKAS...48..113O, Giannini_2019A&A...631A..44G}. The different density regimes likely reflect the distinct regions traced by [S II] and [O I], with the former probing lower-density extended, shock-excited gas and the latter denser regions closer to the wind-launching or shock-cooling zones \citep{Hartigan_1995ApJ...452..736H}. 

\subsection{Implications for the Nature of V1180 Cas}
Taken together, the observational results suggest that V1180 Cas is an actively accreting, moderately inclined EXor-type YSO hosting a structured, multi-component outflow. The simultaneous presence of strong accretion signatures and stratified outflow kinematics is consistent with a magnetically regulated star--disk interaction, in which material is accreted onto the stellar surface while mass is simultaneously ejected through disk winds and jets.

The disk inclination derived in this work ($i_{disk}\approx50^\circ$) lies within the range expected for UXor-like variability, lending further support to the interpretation that both variable circumstellar extinction and episodic accretion contribute to the observed photometric behavior of V1180 Cas \citep{Chand_2026ApJ...998..204C}. The observed kinematic properties and the inferred physical conditions of the emitting gas are also consistent with a scenario in which the inner circumstellar disk plays an important role in launching and shaping the outflow.

When considered together with previous photometric and spectroscopic studies of this source \citep{Kun_2011, Antoniucci_2014A&A...565L...7A, Mutafov_2022RAA....22l5014M, Chand_2026ApJ...998..204C}, our results provide additional evidence that accretion, circumstellar extinction, and mass-loss are interconnected processes governing the variability and circumstellar environment of V1180 Cas.

%%%%%%%%%%%%%%%%%%%%%%%%%%%%%%%%%%%%%%%%%%%%%%%%%%%%%%%%%%%%%%%%%%%%%%%%%%%%%%
\section{Summary and Conclusion}
\label{sec:summary}
In this work, we have presented a detailed analysis of a high-resolution optical spectrum of V1180 Cas to investigate its kinematic properties, accretion activity, and outflow characteristics. The main results of this study are summarized below:
 \begin{enumerate}
\item The optical spectrum is dominated by strong emission lines, including H$\alpha$, Ca II IRT, He I, O I, and numerous Fe I and Fe II transitions, along with a limited number of forbidden lines such as [O I] and [S II], indicating active accretion and outflow processes.
\item A weak Li I $\lambda$6708 absorption feature is detected, confirming the youth of the source. Using this line along with selected Fe I emission lines, we report, for the first time, a RV of $-16 \pm 3$ km\,s$^{-1}$ for V1180 Cas, corresponding to $V_{\rm LSR}\sim-13$ km s$^{-1}$. This value is consistent with the systemic velocity of the parent molecular cloud L1340 ($V_{\rm LSR}=-14.13$ to $-15.92$ km s$^{-1}$; \citealt{Kun_2003A&A...398..169K}), supporting the physical association of V1180 Cas with the cloud.
\item The H$\alpha$ and H$\beta$ line profiles exhibit asymmetric shapes with excess emission in the blue wing and blueshifted absorption components, indicative of outflowing material along the line of sight.
\item The He I $\lambda$5876 line shows a two-component structure: a NC near the systemic velocity tracing post-shock accretion regions, and a broad, slightly blueshifted component (BC) associated with magnetospheric flows and/or inner disk winds.
\item The [O I] $\lambda$6300 line profile is decomposed into a LVC ($\sim -40$ km s$^{-1}$, FWHM $\sim 64$ km s$^{-1}$) and a HVC ($\sim -76$ km s$^{-1}$, FWHM $\sim 170$ km s$^{-1}$), tracing a stratified outflow consisting of a slow disk wind and a fast collimated jet.
\item Using the forbidden [O I] $\lambda$6300 line, we estimate, for the first time, a disk inclination angle of $i_{\mathrm{disk}} \approx 50^\circ$, suggesting a moderately inclined system geometry.
\item The mass accretion rates derived from H$\alpha$, H$\beta$, He I, O I, and the Ca II IRT lines span $\dot{M}_{acc}\sim(7.7$--$23.6)\times10^{-7}$ $M_{\odot}$ yr$^{-1}$, consistent with previous estimates during bright states and typical of EXor-type objects.
\item The jet mass-loss rate estimated from the HVC of the [O I] $\lambda$6300 line is $\sim 2.6 \times 10^{-8}$ $M_{\odot}$ yr$^{-1}$, and the wind mass-loss rate from the LVC is $\sim1.8\times10^{-8}$ $M_{\odot}$ yr$^{-1}$, yielding $\dot{M}_{jet}/\dot{M}_{acc} \approx 0.01$--$0.03$, at the lower end of the range reported for Class II YSOs but consistent with magnetically driven jet-launching models. However, the calculated mass-loss rates for the jet and wind are subject to substantial systematic uncertainties, particularly due to the assumed emitting-region size and filling factor.
\item Forbidden-line diagnostics indicate a range of physical conditions in the outflow, with electron densities $\sim 10^{3}$ and $10^{6}$ cm$^{-3}$ with characteristic temperatures of $\sim 10^{4}$ K, supporting a multi-component origin involving both dense disk winds and shock-excited jet material. However, these estimates are subject to uncertainties associated with the adopted diagnostics and assumptions about the physical conditions of the emitting regions.
 \end{enumerate}

\par Overall, this study provides a comprehensive view of the inner accretion--ejection environment of V1180 Cas, highlighting the complex and interconnected processes that govern the evolution of YSOs.
\\

\section*{Acknowledgments}
We thank the anonymous reviewer for the comments and suggestions that enhanced the quality of the paper.
This research has made use of the Keck Observatory Archive (KOA), which is operated by the W. M. Keck Observatory and the NASA Exoplanet Science Institute (NExScI), under contract with the National Aeronautics and Space Administration.
K.S. and J.P.N were supported in part by a generous donation (from the Murty Trust) aimed at enabling advances in astrophysics through the use of machine learning. The Murty Trust, an initiative of the Murty Foundation, is a not-for-profit organization dedicated to the preservation and celebration of culture, science, and knowledge
of systems born out of India. The Murty Trust is headed by
Mrs. Sudha Murty and Mr. Rohan Murty. 
J.P.N., D.K.O. and J.D. acknowledge the support of the Department of Atomic Energy, Government of India, under project identification No. RTI 4012.
M.C. acknowledges the financial support of DST-INSPIRE (No.: DST/INSPIRE Fellowship/2022/IF220587).
A.V. acknowledges the financial support of DST-INSPIRE (No.: DST/INSPIRE Fellowship/2019/IF190550).

%%%%%%%%%%%%%%%%%%%% REFERENCES %%%%%%%%%%%%%%%%%%

\bibliography{thebibliography}

@INPROCEEDINGS{1993AAS...183.1101V,
       author = {{Vogt}, Steven S. and {Keane}, Michael J.},
        title = "{HIRES: The High Resolution Echelle Spectrometer on the Keck Telescope}",
    booktitle = {American Astronomical Society Meeting Abstracts},
         year = 1993,
       series = {American Astronomical Society Meeting Abstracts},
       volume = {183},
        month = dec,
          eid = {11.01},
        pages = {11.01},
       adsurl = {https://ui.adsabs.harvard.edu/abs/1993AAS...183.1101V}
}

@article{Kun_2011,
doi = {10.1088/2041-8205/733/1/L8},
url = {https://dx.doi.org/10.1088/2041-8205/733/1/L8},
year = {2011},
month = {apr},
publisher = {The American Astronomical Society},
volume = {733},
number = {1},
pages = {L8},
author = {M. Kun and E. Szegedi-Elek and A. Moór and P. Ábrahám and J. A. Acosta-Pulido and D. Apai and J. Kelemen and A. Pál and M. Rácz and Zs. Regály and R. Szakáts and N. Szalai and A. Szing},
title = {A PECULIAR YOUNG ERUPTIVE STAR IN THE DARK CLOUD LYNDS 1340},
journal = {The Astrophysical Journal Letters}
}

@ARTICLE{Antoniucci_2014A&A...565L...7A,
       author = {{Antoniucci}, S. and {Arkharov}, A.~A. and {Di Paola}, A. and {Giannini}, T. and {Harutyunyan}, A. and {Kopatskaya}, E.~N. and {Larionov}, V.~M. and {Li Causi}, G. and {Lorenzetti}, D. and {Morozova}, D. and {Nisini}, B. and {Vitali}, F.},
        title = "{Recent outburst of the young star V1180 Cassiopeiae}",
      journal = {\aap},
         year = 2014,
        month = may,
       volume = {565},
          eid = {L7},
        pages = {L7},
          doi = {10.1051/0004-6361/201423962},
archivePrefix = {arXiv},
       eprint = {1405.0612},
 primaryClass = {astro-ph.SR},
       adsurl = {https://ui.adsabs.harvard.edu/abs/2014A&A...565L...7A}
}

@ARTICLE{Mutafov_2022RAA....22l5014M,
       author = {{Mutafov}, Asen and {Semkov}, Evgeni and {Peneva}, Stoyanka and {Ibryamov}, Sunay},
        title = "{Long-term Photometric Study of the Pre-main Sequence Star V1180 Cas}",
      journal = {Research in Astronomy and Astrophysics},
         year = 2022,
        month = dec,
       volume = {22},
       number = {12},
          eid = {125014},
        pages = {125014},
          doi = {10.1088/1674-4527/ac9af0},
archivePrefix = {arXiv},
       eprint = {2210.09660},
 primaryClass = {astro-ph.SR},
       adsurl = {https://ui.adsabs.harvard.edu/abs/2022RAA....22l5014M}
}

@ARTICLE{Kun_1994A&A...292..249K,
       author = {{Kun}, M. and {Obayashi}, A. and {Sato}, F. and {Yonekura}, Y. and {Fukui}, Y. and {Balazs}, L.~G. and {Abraham}, P. and {Szabados}, L. and {Kelemen}, J.},
        title = "{Study of L1340: a star-forming cloud in Cassiopeia.}",
      journal = {\aap},
         year = 1994,
        month = dec,
       volume = {292},
        pages = {249-260},
       adsurl = {https://ui.adsabs.harvard.edu/abs/1994A&A...292..249K}
}

@ARTICLE{Abraham_2018ApJ...853...28A,
       author = {{{\'A}brah{\'a}m}, P. and {K{\'o}sp{\'a}l}, {\'A}. and {Kun}, M. and {Feh{\'e}r}, O. and {Zsidi}, G. and {Acosta-Pulido}, J.~A. and {Carnerero}, M.~I. and {Garc{\'\i}a-{\'A}lvarez}, D. and {Mo{\'o}r}, A. and {Cseh}, B. and {Hajdu}, G. and {Hanyecz}, O. and {Kelemen}, J. and {Kriskovics}, L. and {Marton}, G. and {Mez{\H{o}}}, Gy. and {Moln{\'a}r}, L. and {Ordasi}, A. and {Rodr{\'\i}guez-Coira}, G. and {S{\'a}rneczky}, K. and {S{\'o}dor}, {\'A}. and {Szak{\'a}ts}, R. and {Szegedi-Elek}, E. and {Szing}, A. and {Farkas-Tak{\'a}cs}, A. and {Vida}, K. and {Vink{\'o}}, J.},
        title = "{An UXor among FUors: Extinction-related Brightness Variations of the Young Eruptive Star V582 Aur}",
      journal = {\apj},
         year = 2018,
        month = jan,
       volume = {853},
       number = {1},
          eid = {28},
        pages = {28},
          doi = {10.3847/1538-4357/aaa242},
archivePrefix = {arXiv},
       eprint = {1712.04968},
 primaryClass = {astro-ph.SR},
       adsurl = {https://ui.adsabs.harvard.edu/abs/2018ApJ...853...28A}
}

@ARTICLE{Natta_2000A&A...364..633N,
       author = {{Natta}, A. and {Whitney}, B.~A.},
        title = "{Models of scattered light in UXORs}",
      journal = {\aap},
         year = 2000,
        month = dec,
       volume = {364},
        pages = {633-640},
       adsurl = {https://ui.adsabs.harvard.edu/abs/2000A&A...364..633N}
}

@INPROCEEDINGS{Pascucci_2023ASPC..534..567P,
       author = {{Pascucci}, I. and {Cabrit}, S. and {Edwards}, S. and {Gorti}, U. and {Gressel}, O. and {Suzuki}, T.~K.},
        title = "{The Role of Disk Winds in the Evolution and Dispersal of Protoplanetary Disks}",
    booktitle = {Protostars and Planets VII},
         year = 2023,
       editor = {{Inutsuka}, S. and {Aikawa}, Y. and {Muto}, T. and {Tomida}, K. and {Tamura}, M.},
       series = {Astronomical Society of the Pacific Conference Series},
       volume = {534},
        month = jul,
        pages = {567},
          doi = {10.48550/arXiv.2203.10068},
archivePrefix = {arXiv},
       eprint = {2203.10068},
 primaryClass = {astro-ph.EP},
       adsurl = {https://ui.adsabs.harvard.edu/abs/2023ASPC..534..567P}
}

@ARTICLE{Alcala_2014A&A...561A...2A,
       author = {{Alcal{\'a}}, J.~M. and {Natta}, A. and {Manara}, C.~F. and {Spezzi}, L. and {Stelzer}, B. and {Frasca}, A. and {Biazzo}, K. and {Covino}, E. and {Randich}, S. and {Rigliaco}, E. and {Testi}, L. and {Comer{\'o}n}, F. and {Cupani}, G. and {D'Elia}, V.},
        title = "{X-shooter spectroscopy of young stellar objects. IV. Accretion in low-mass stars and substellar objects in Lupus}",
      journal = {\aap},
         year = 2014,
        month = jan,
       volume = {561},
          eid = {A2},
        pages = {A2},
          doi = {10.1051/0004-6361/201322254},
archivePrefix = {arXiv},
       eprint = {1310.2069},
 primaryClass = {astro-ph.SR},
       adsurl = {https://ui.adsabs.harvard.edu/abs/2014A&A...561A...2A}
}

@ARTICLE{White_2003ApJ...582.1109W,
       author = {{White}, Russel J. and {Basri}, Gibor},
        title = "{Very Low Mass Stars and Brown Dwarfs in Taurus-Auriga}",
      journal = {\apj},
         year = 2003,
        month = jan,
       volume = {582},
       number = {2},
        pages = {1109-1122},
          doi = {10.1086/344673},
archivePrefix = {arXiv},
       eprint = {astro-ph/0209164},
 primaryClass = {astro-ph},
       adsurl = {https://ui.adsabs.harvard.edu/abs/2003ApJ...582.1109W}
}

@ARTICLE{Muzerolle_1998AJ....116..455M,
       author = {{Muzerolle}, James and {Hartmann}, Lee and {Calvet}, Nuria},
        title = "{Emission-Line Diagnostics of T Tauri Magnetospheric Accretion. I. Line Profile Observations}",
      journal = {\aj},
         year = 1998,
        month = jul,
       volume = {116},
       number = {1},
        pages = {455-468},
          doi = {10.1086/300428},
       adsurl = {https://ui.adsabs.harvard.edu/abs/1998AJ....116..455M}
}

@ARTICLE{Ninan_2015ApJ...815....4N,
       author = {{Ninan}, J.~P. and {Ojha}, D.~K. and {Baug}, T. and {Bhatt}, B.~C. and {Mohan}, V. and {Ghosh}, S.~K. and {Men'shchikov}, A. and {Anupama}, G.~C. and {Tamura}, M. and {Henning}, Th.},
        title = "{V899 Mon: An Outbursting Protostar with a Peculiar Light Curve, and Its Transition Phases}",
      journal = {\apj},
         year = 2015,
        month = dec,
       volume = {815},
       number = {1},
          eid = {4},
        pages = {4},
          doi = {10.1088/0004-637X/815/1/4},
archivePrefix = {arXiv},
       eprint = {1510.07116},
 primaryClass = {astro-ph.SR},
       adsurl = {https://ui.adsabs.harvard.edu/abs/2015ApJ...815....4N}
}

@ARTICLE{Alcala_2017A&A...600A..20A,
       author = {{Alcal{\'a}}, J.~M. and {Manara}, C.~F. and {Natta}, A. and {Frasca}, A. and {Testi}, L. and {Nisini}, B. and {Stelzer}, B. and {Williams}, J.~P. and {Antoniucci}, S. and {Biazzo}, K. and {Covino}, E. and {Esposito}, M. and {Getman}, F. and {Rigliaco}, E.},
        title = "{X-shooter spectroscopy of young stellar objects in Lupus. Accretion properties of class II and transitional objects}",
      journal = {\aap},
         year = 2017,
        month = apr,
       volume = {600},
          eid = {A20},
        pages = {A20},
          doi = {10.1051/0004-6361/201629929},
archivePrefix = {arXiv},
       eprint = {1612.07054},
 primaryClass = {astro-ph.SR},
       adsurl = {https://ui.adsabs.harvard.edu/abs/2017A&A...600A..20A}
}

@ARTICLE{Kun_2016ApJ...822...79K,
       author = {{Kun}, M{\'a}ria and {Mo{\'o}r}, Attila and {Szegedi-Elek}, Elza and {Reipurth}, Bo},
        title = "{The Intermediate-mass Star-forming Region Lynds 1340. An Optical View}",
      journal = {\apj},
         year = 2016,
        month = may,
       volume = {822},
       number = {2},
          eid = {79},
        pages = {79},
          doi = {10.3847/0004-637X/822/2/79},
archivePrefix = {arXiv},
       eprint = {1602.08120},
 primaryClass = {astro-ph.SR},
       adsurl = {https://ui.adsabs.harvard.edu/abs/2016ApJ...822...79K}
}

@ARTICLE{Hartigan_1995ApJ...452..736H,
       author = {{Hartigan}, Patrick and {Edwards}, Suzan and {Ghandour}, Louma},
        title = "{Disk Accretion and Mass Loss from Young Stars}",
      journal = {\apj},
         year = 1995,
        month = oct,
       volume = {452},
        pages = {736},
          doi = {10.1086/176344},
       adsurl = {https://ui.adsabs.harvard.edu/abs/1995ApJ...452..736H}
}

@ARTICLE{Appenzeller_2013A&A...558A..83A,
       author = {{Appenzeller}, Immo and {Bertout}, Claude},
        title = "{Inclination effects in T Tauri star spectra}",
      journal = {\aap},
         year = 2013,
        month = oct,
       volume = {558},
          eid = {A83},
        pages = {A83},
          doi = {10.1051/0004-6361/201322160},
archivePrefix = {arXiv},
       eprint = {1308.3095},
 primaryClass = {astro-ph.SR},
       adsurl = {https://ui.adsabs.harvard.edu/abs/2013A&A...558A..83A}
}

@INPROCEEDINGS{Fischer_2023ASPC..534..355F,
       author = {{Fischer}, W.~J. and {Hillenbrand}, L.~A. and {Herczeg}, G.~J. and {Johnstone}, D. and {Kospal}, A. and {Dunham}, M.~M.},
        title = "{Accretion Variability as a Guide to Stellar Mass Assembly}",
    booktitle = {Protostars and Planets VII},
         year = 2023,
       editor = {{Inutsuka}, S. and {Aikawa}, Y. and {Muto}, T. and {Tomida}, K. and {Tamura}, M.},
       series = {Astronomical Society of the Pacific Conference Series},
       volume = {534},
        month = jul,
        pages = {355},
          doi = {10.48550/arXiv.2203.11257},
archivePrefix = {arXiv},
       eprint = {2203.11257},
 primaryClass = {astro-ph.SR},
       adsurl = {https://ui.adsabs.harvard.edu/abs/2023ASPC..534..355F}
}

@ARTICLE{Herbig_2008AJ....135..637H,
       author = {{Herbig}, G.~H.},
        title = "{History and Spectroscopy of EXor Candidates}",
      journal = {\aj},
         year = 2008,
        month = feb,
       volume = {135},
       number = {2},
        pages = {637-648},
          doi = {10.1088/0004-6256/135/2/637},
       adsurl = {https://ui.adsabs.harvard.edu/abs/2008AJ....135..637H}
}

@ARTICLE{Herbig_1977ApJ...217..693H,
       author = {{Herbig}, G.~H.},
        title = "{Eruptive phenomena in early stellar evolution.}",
      journal = {\apj},
         year = 1977,
        month = nov,
       volume = {217},
        pages = {693-715},
          doi = {10.1086/155615},
       adsurl = {https://ui.adsabs.harvard.edu/abs/1977ApJ...217..693H}
}

@ARTICLE{Herbst_1994AJ....108.1906H,
       author = {{Herbst}, W. and {Herbst}, D.~K. and {Grossman}, E.~J. and {Weinstein}, D.},
        title = "{Catalogue of UBVRI Photometry of T Tauri Stars and Analysis of the Causes of Their Variability}",
      journal = {\aj},
         year = 1994,
        month = nov,
       volume = {108},
        pages = {1906},
          doi = {10.1086/117204},
       adsurl = {https://ui.adsabs.harvard.edu/abs/1994AJ....108.1906H}
}

@ARTICLE{Sicilia_2012A&A...544A..93S,
       author = {{Sicilia-Aguilar}, A. and {K{\'o}sp{\'a}l}, {\'A}. and {Setiawan}, J. and {{\'A}brah{\'a}m}, P. and {Dullemond}, C. and {Eiroa}, C. and {Goto}, M. and {Henning}, Th. and {Juh{\'a}sz}, A.},
        title = "{Optical spectroscopy of EX Lupi during quiescence and outburst. Infall, wind, and dynamics in the accretion flow}",
      journal = {\aap},
         year = 2012,
        month = aug,
       volume = {544},
          eid = {A93},
        pages = {A93},
          doi = {10.1051/0004-6361/201118555},
archivePrefix = {arXiv},
       eprint = {1206.3081},
 primaryClass = {astro-ph.SR},
       adsurl = {https://ui.adsabs.harvard.edu/abs/2012A&A...544A..93S}
}

@ARTICLE{Miera_2023A&A...678A..88C,
       author = {{Cruz-S{\'a}enz de Miera}, F. and {K{\'o}sp{\'a}l}, {\'A}. and {Abrah{\'a}m}, P. and {Claes}, R.~A.~B. and {Manara}, C.~F. and {Wendeborn}, J. and {Fiorellino}, E. and {Giannini}, T. and {Nisini}, B. and {Sicilia-Aguilar}, A. and {Campbell-White}, J. and {Alcal{\'a}}, J.~M. and {Banzatti}, A. and {Szab{\'o}}, Zs. M. and {Lykou}, F. and {Antoniucci}, S. and {Varga}, J. and {Siwak}, M. and {Park}, S. and {Nagy}, Zs. and {Kun}, M.},
        title = "{Brightness and mass accretion rate evolution during the 2022 burst of EX Lupi}",
      journal = {\aap},
         year = 2023,
        month = oct,
       volume = {678},
          eid = {A88},
        pages = {A88},
          doi = {10.1051/0004-6361/202347063},
archivePrefix = {arXiv},
       eprint = {2308.02849},
 primaryClass = {astro-ph.SR},
       adsurl = {https://ui.adsabs.harvard.edu/abs/2023A&A...678A..88C}
}

@ARTICLE{Natta_2004A&A...424..603N,
       author = {{Natta}, A. and {Testi}, L. and {Muzerolle}, J. and {Randich}, S. and {Comer{\'o}n}, F. and {Persi}, P.},
        title = "{Accretion in brown dwarfs: An infrared view}",
      journal = {\aap},
         year = 2004,
        month = sep,
       volume = {424},
        pages = {603-612},
          doi = {10.1051/0004-6361:20040356},
archivePrefix = {arXiv},
       eprint = {astro-ph/0406106},
 primaryClass = {astro-ph},
       adsurl = {https://ui.adsabs.harvard.edu/abs/2004A&A...424..603N}
}

@ARTICLE{Fang_2013ApJS..207....5F,
       author = {{Fang}, Min and {Kim}, Jinyoung Serena and {van Boekel}, Roy and {Sicilia-Aguilar}, Aurora and {Henning}, Thomas and {Flaherty}, Kevin},
        title = "{Young Stellar Objects in Lynds 1641: Disks, Accretion, and Star Formation History}",
      journal = {\apjs},
         year = 2013,
        month = jul,
       volume = {207},
       number = {1},
          eid = {5},
        pages = {5},
          doi = {10.1088/0067-0049/207/1/5},
archivePrefix = {arXiv},
       eprint = {1304.7777},
 primaryClass = {astro-ph.SR},
       adsurl = {https://ui.adsabs.harvard.edu/abs/2013ApJS..207....5F}
}

@ARTICLE{Hamann_1992ApJS...82..247H,
       author = {{Hamann}, Fred and {Persson}, S.~E.},
        title = "{Emission-Line Studies of Young Stars. I. The T Tauri Stars}",
      journal = {\apjs},
         year = 1992,
        month = sep,
       volume = {82},
        pages = {247},
          doi = {10.1086/191715},
       adsurl = {https://ui.adsabs.harvard.edu/abs/1992ApJS...82..247H}
}

@ARTICLE{Singh_2024ApJ...968...88S,
       author = {{Singh}, Koshvendra and {Ninan}, Joe P. and {Romanova}, Marina M. and {Buckley}, David A.~H. and {Ojha}, Devendra K. and {Ghosh}, Arpan and {Monson}, Andrew and {Schramm}, Malte and {Sharma}, Saurabh and {Reichart}, Daniel E. and {Mikolajewska}, Joanna and {Beamin}, Juan Carlos and {Borissova}, Jura and {Ivanov}, Valentin D. and {Kouprianov}, Vladimir V. and {Hambsch}, Franz-Josef and {Pearce}, Andrew},
        title = "{Accretion Funnel Reconfiguration during an Outburst in a Young Stellar Object: EX Lupi}",
      journal = {\apj},
         year = 2024,
        month = jun,
       volume = {968},
       number = {2},
          eid = {88},
        pages = {88},
          doi = {10.3847/1538-4357/ad4099},
archivePrefix = {arXiv},
       eprint = {2404.05420},
 primaryClass = {astro-ph.SR},
       adsurl = {https://ui.adsabs.harvard.edu/abs/2024ApJ...968...88S}
}

@ARTICLE{pyneb_2015A&A...573A..42L,
       author = {{Luridiana}, V. and {Morisset}, C. and {Shaw}, R.~A.},
        title = "{PyNeb: a new tool for analyzing emission lines. I. Code description and validation of results}",
      journal = {\aap},
         year = 2015,
        month = jan,
       volume = {573},
          eid = {A42},
        pages = {A42},
          doi = {10.1051/0004-6361/201323152},
archivePrefix = {arXiv},
       eprint = {1410.6662},
 primaryClass = {astro-ph.IM},
       adsurl = {https://ui.adsabs.harvard.edu/abs/2015A&A...573A..42L}
}

@BOOK{Osterbrock_2006agna.book.....O,
       author = {{Osterbrock}, Donald E. and {Ferland}, Gary J.},
        title = "{Astrophysics of gaseous nebulae and active galactic nuclei}",
         year = 2006,
       adsurl = {https://ui.adsabs.harvard.edu/abs/2006agna.book.....O}
}

@ARTICLE{Fang_2018ApJ...868...28F,
       author = {{Fang}, Min and {Pascucci}, Ilaria and {Edwards}, Suzan and {Gorti}, Uma and {Banzatti}, Andrea and {Flock}, Mario and {Hartigan}, Patrick and {Herczeg}, Gregory J. and {Dupree}, Andrea K.},
        title = "{A New Look at T Tauri Star Forbidden Lines: MHD-driven Winds from the Inner Disk}",
      journal = {\apj},
         year = 2018,
        month = nov,
       volume = {868},
       number = {1},
          eid = {28},
        pages = {28},
          doi = {10.3847/1538-4357/aae780},
archivePrefix = {arXiv},
       eprint = {1810.03366},
 primaryClass = {astro-ph.SR},
       adsurl = {https://ui.adsabs.harvard.edu/abs/2018ApJ...868...28F}
}

@ARTICLE{Natta_2014A&A...569A...5N,
       author = {{Natta}, A. and {Testi}, L. and {Alcal{\'a}}, J.~M. and {Rigliaco}, E. and {Covino}, E. and {Stelzer}, B. and {D'Elia}, V.},
        title = "{X-shooter spectroscopy of young stellar objects. V. Slow winds in T Tauri stars}",
      journal = {\aap},
         year = 2014,
        month = sep,
       volume = {569},
          eid = {A5},
        pages = {A5},
          doi = {10.1051/0004-6361/201424136},
archivePrefix = {arXiv},
       eprint = {1406.5630},
 primaryClass = {astro-ph.SR},
       adsurl = {https://ui.adsabs.harvard.edu/abs/2014A&A...569A...5N}
}

@ARTICLE{Sicilia_2020A&A...643A..29S,
       author = {{Sicilia-Aguilar}, A. and {Bouvier}, J. and {Dougados}, C. and {Grankin}, K. and {Donati}, J.~F.},
        title = "{Reading between the lines. Disk emission, wind, and accretion during the Z CMa NW outburst}",
      journal = {\aap},
         year = 2020,
        month = nov,
       volume = {643},
          eid = {A29},
        pages = {A29},
          doi = {10.1051/0004-6361/202038489},
archivePrefix = {arXiv},
       eprint = {2008.13678},
 primaryClass = {astro-ph.SR},
       adsurl = {https://ui.adsabs.harvard.edu/abs/2020A&A...643A..29S}
}

@ARTICLE{Chand_2026ApJ...998..204C,
       author = {{Chand}, Tarak and {Sharma}, Saurabh and {Singh}, Koshvendra and {Ninan}, Joe P. and {Ghosh}, Arpan and {Ojha}, Devendra K. and {Baug}, Tapas and {Sahu}, D.~K. and {Bhatt}, Bhuwan C. and {Kumar}, Pramod and {Yadav}, Ram K. and {Panwar}, Neelam and {Verma}, Aayushi and {Kaur}, Harmeen and {Mamta} and {Chakraborty}, Manojit and {Gokhe}, Kartik and {Singh}, Ajay Kumar},
        title = "{Unveiling the Dual Nature of V1180 Cas: UXor-like Dips and EXor-like Bursts across a Decade}",
      journal = {\apj},
         year = 2026,
        month = feb,
       volume = {998},
       number = {2},
          eid = {204},
        pages = {204},
          doi = {10.3847/1538-4357/ae3180},
archivePrefix = {arXiv},
       eprint = {2512.20085},
 primaryClass = {astro-ph.SR},
       adsurl = {https://ui.adsabs.harvard.edu/abs/2026ApJ...998..204C}
}

@ARTICLE{Bacciotti_1999A&A...350..917B,
       author = {{Bacciotti}, Francesca and {Eisl{\"o}ffel}, Jochen and {Ray}, Thomas P.},
        title = "{The physical properties of the HH 30 jet from HST and ground-based data}",
      journal = {\aap},
         year = 1999,
        month = oct,
       volume = {350},
        pages = {917-927},
       adsurl = {https://ui.adsabs.harvard.edu/abs/1999A&A...350..917B}
}

@ARTICLE{Agra-Amboage_2011A&A...532A..59A,
       author = {{Agra-Amboage}, V. and {Dougados}, C. and {Cabrit}, S. and {Reunanen}, J.},
        title = "{Sub-arcsecond [Fe ii] spectro-imaging of the DG Tauri jet. Periodic bubbles and a dusty disk wind?}",
      journal = {\aap},
         year = 2011,
        month = aug,
       volume = {532},
          eid = {A59},
        pages = {A59},
          doi = {10.1051/0004-6361/201015886},
archivePrefix = {arXiv},
       eprint = {1106.2690},
 primaryClass = {astro-ph.SR},
       adsurl = {https://ui.adsabs.harvard.edu/abs/2011A&A...532A..59A}
}

@ARTICLE{Ferreira_2006A&A...453..785F,
       author = {{Ferreira}, J. and {Dougados}, C. and {Cabrit}, S.},
        title = "{Which jet launching mechanism(s) in T Tauri stars?}",
      journal = {\aap},
         year = 2006,
        month = jul,
       volume = {453},
       number = {3},
        pages = {785-796},
          doi = {10.1051/0004-6361:20054231},
archivePrefix = {arXiv},
       eprint = {astro-ph/0604053},
 primaryClass = {astro-ph},
       adsurl = {https://ui.adsabs.harvard.edu/abs/2006A&A...453..785F}
}

@ARTICLE{Cardelli_1989ApJ...345..245C,
       author = {{Cardelli}, Jason A. and {Clayton}, Geoffrey C. and {Mathis}, John S.},
        title = "{The Relationship between Infrared, Optical, and Ultraviolet Extinction}",
      journal = {\apj},
         year = 1989,
        month = oct,
       volume = {345},
        pages = {245},
          doi = {10.1086/167900},
       adsurl = {https://ui.adsabs.harvard.edu/abs/1989ApJ...345..245C}
}

@ARTICLE{Hartmann_2016ARA&A..54..135H,
       author = {{Hartmann}, Lee and {Herczeg}, Gregory and {Calvet}, Nuria},
        title = "{Accretion onto Pre-Main-Sequence Stars}",
      journal = {\araa},
         year = 2016,
        month = sep,
       volume = {54},
        pages = {135-180},
          doi = {10.1146/annurev-astro-081915-023347},
       adsurl = {https://ui.adsabs.harvard.edu/abs/2016ARA&A..54..135H}
}

@INPROCEEDINGS{Ray_2007prpl.conf..231R,
       author = {{Ray}, T. and {Dougados}, C. and {Bacciotti}, F. and {Eisl{\"o}ffel}, J. and {Chrysostomou}, A.},
        title = "{Toward Resolving the Outflow Engine: An Observational Perspective}",
    booktitle = {Protostars and Planets V},
         year = 2007,
       editor = {{Reipurth}, Bo and {Jewitt}, David and {Keil}, Klaus},
        month = jan,
        pages = {231},
          doi = {10.48550/arXiv.astro-ph/0605597},
archivePrefix = {arXiv},
       eprint = {astro-ph/0605597},
 primaryClass = {astro-ph},
       adsurl = {https://ui.adsabs.harvard.edu/abs/2007prpl.conf..231R}
}

@ARTICLE{Nisini_2018A&A...609A..87N,
       author = {{Nisini}, B. and {Antoniucci}, S. and {Alcal{\'a}}, J.~M. and {Giannini}, T. and {Manara}, C.~F. and {Natta}, A. and {Fedele}, D. and {Biazzo}, K.},
        title = "{Connection between jets, winds and accretion in T Tauri stars. The X-shooter view}",
      journal = {\aap},
         year = 2018,
        month = jan,
       volume = {609},
          eid = {A87},
        pages = {A87},
          doi = {10.1051/0004-6361/201730834},
archivePrefix = {arXiv},
       eprint = {1710.05587},
 primaryClass = {astro-ph.SR},
       adsurl = {https://ui.adsabs.harvard.edu/abs/2018A&A...609A..87N}
}

@ARTICLE{Kwan_1995ApJ...454..382K,
       author = {{Kwan}, John and {Tademaru}, Eugene},
        title = "{Disk Winds from T Tauri Stars}",
      journal = {\apj},
         year = 1995,
        month = nov,
       volume = {454},
        pages = {382},
          doi = {10.1086/176489},
       adsurl = {https://ui.adsabs.harvard.edu/abs/1995ApJ...454..382K}
}

@ARTICLE{Cabrit_1999A&A...343L..61C,
       author = {{Cabrit}, S. and {Ferreira}, J. and {Raga}, A.~C.},
        title = "{Forbidden lines from T Tauri disk winds. I. High magnetic torque models}",
      journal = {\aap},
         year = 1999,
        month = mar,
       volume = {343},
        pages = {L61-L64},
       adsurl = {https://ui.adsabs.harvard.edu/abs/1999A&A...343L..61C}
}

@ARTICLE{McGinnis_2018A&A...620A..87M,
       author = {{McGinnis}, P. and {Dougados}, C. and {Alencar}, S.~H.~P. and {Bouvier}, J. and {Cabrit}, S.},
        title = "{Ejection processes in the young open cluster NGC 2264. A study of the [OI]{\ensuremath{\lambda}}6300{\r{A}} emission line}",
      journal = {\aap},
         year = 2018,
        month = nov,
       volume = {620},
          eid = {A87},
        pages = {A87},
          doi = {10.1051/0004-6361/201731629},
archivePrefix = {arXiv},
       eprint = {1803.10287},
 primaryClass = {astro-ph.SR},
       adsurl = {https://ui.adsabs.harvard.edu/abs/2018A&A...620A..87M}
}

@ARTICLE{Sestito_2008A&A...487..965S,
       author = {{Sestito}, P. and {Palla}, F. and {Randich}, S.},
        title = "{An assessment of Li abundances in weak-lined and classical T Tauri stars of the Taurus-Auriga association}",
      journal = {\aap},
         year = 2008,
        month = sep,
       volume = {487},
       number = {3},
        pages = {965-973},
          doi = {10.1051/0004-6361:200810081},
archivePrefix = {arXiv},
       eprint = {0807.0302},
 primaryClass = {astro-ph},
       adsurl = {https://ui.adsabs.harvard.edu/abs/2008A&A...487..965S}
}

@ARTICLE{Lima_2010A&A...522A.104L,
       author = {{Lima}, G.~H.~R.~A. and {Alencar}, S.~H.~P. and {Calvet}, N. and {Hartmann}, L. and {Muzerolle}, J.},
        title = "{Modeling the H{\ensuremath{\alpha}} line emission around classical T Tauri stars using magnetospheric accretion and disk wind models}",
      journal = {\aap},
         year = 2010,
        month = nov,
       volume = {522},
          eid = {A104},
        pages = {A104},
          doi = {10.1051/0004-6361/201014490},
archivePrefix = {arXiv},
       eprint = {1007.3976},
 primaryClass = {astro-ph.SR},
       adsurl = {https://ui.adsabs.harvard.edu/abs/2010A&A...522A.104L}
}

@ARTICLE{Kurosawa_2012MNRAS.426.2901K,
       author = {{Kurosawa}, Ryuichi and {Romanova}, M.~M.},
        title = "{Line formation in the inner winds of classical T Tauri stars: testing the conical-shell wind solution}",
      journal = {\mnras},
         year = 2012,
        month = nov,
       volume = {426},
       number = {4},
        pages = {2901-2916},
          doi = {10.1111/j.1365-2966.2012.21853.x},
archivePrefix = {arXiv},
       eprint = {1203.5154},
 primaryClass = {astro-ph.SR},
       adsurl = {https://ui.adsabs.harvard.edu/abs/2012MNRAS.426.2901K}
}

@ARTICLE{Herbig_1980ApJ...242..628H,
       author = {{Herbig}, G.~H. and {Soderblom}, D.~R.},
        title = "{Observations and interpretation of the near-infrared line spectra of T Tau stars.}",
      journal = {\apj},
         year = 1980,
        month = dec,
       volume = {242},
        pages = {628-637},
          doi = {10.1086/158499},
       adsurl = {https://ui.adsabs.harvard.edu/abs/1980ApJ...242..628H}
}

@article{Azevedo_2006,
	author = {{Azevedo, R.} and {Calvet, N.} and {Hartmann, L.} and {Folha, D. F. M.} and {Gameiro, F.} and {Muzerolle, J.}},
	title = {Calcium II~infrared triplet line models in classical T Tauri stars},
	DOI= "10.1051/0004-6361:20054315",
	url= "https://doi.org/10.1051/0004-6361:20054315",
	journal = {A\&A},
	year = 2006,
	volume = 456,
	number = 1,
	pages = "225-234",
}

@ARTICLE{Hodapp_2019AJ....158..241H,
       author = {{Hodapp}, Klaus W. and {Reipurth}, Bo and {Pettersson}, Bertil and {Tonry}, John and {Denneau}, Larry and {Vallely}, Patrick J. and {Shappee}, Benjamin J. and {Armstrong}, James D. and {Connelley}, Michael S. and {Kochanek}, C.~S. and {Fausnaugh}, Michael and {Chini}, Rolf and {Haas}, Martin and {Sobrino Figaredo}, Catalina},
        title = "{The New EXor Outburst of ESO-H{\ensuremath{\alpha}} 99 Observed by Gaia ATLAS and TESS}",
      journal = {\aj},
         year = 2019,
        month = dec,
       volume = {158},
       number = {6},
          eid = {241},
        pages = {241},
          doi = {10.3847/1538-3881/ab471a},
archivePrefix = {arXiv},
       eprint = {1909.10636},
 primaryClass = {astro-ph.SR},
       adsurl = {https://ui.adsabs.harvard.edu/abs/2019AJ....158..241H}
}

@ARTICLE{Hodapp_2020AJ....160..164H,
       author = {{Hodapp}, Klaus W. and {Denneau}, Larry and {Tucker}, Michael and {Shappee}, Benjamin J. and {Huber}, Mark E. and {Payne}, Anna V. and {Do}, Aaron and {Lin}, Chien-Cheng and {Connelley}, Michael S. and {Varricatt}, Watson P. and {Tonry}, John and {Chambers}, Kenneth and {Magnier}, Eugene},
        title = "{The Outburst of the Young Star Gaia19bey}",
      journal = {\aj},
         year = 2020,
        month = oct,
       volume = {160},
       number = {4},
          eid = {164},
        pages = {164},
          doi = {10.3847/1538-3881/abad96},
archivePrefix = {arXiv},
       eprint = {2008.03396},
 primaryClass = {astro-ph.SR},
       adsurl = {https://ui.adsabs.harvard.edu/abs/2020AJ....160..164H}
}

@ARTICLE{Ninan_2013ApJ...778..116N,
       author = {{Ninan}, J.~P. and {Ojha}, D.~K. and {Bhatt}, B.~C. and {Ghosh}, S.~K. and {Mohan}, V. and {Mallick}, K.~K. and {Tamura}, M. and {Henning}, Th.},
        title = "{Reappearance of McNeil's Nebula (V1647 Orionis) and its Outburst Environment}",
      journal = {\apj},
         year = 2013,
        month = dec,
       volume = {778},
       number = {2},
          eid = {116},
        pages = {116},
          doi = {10.1088/0004-637X/778/2/116},
archivePrefix = {arXiv},
       eprint = {1309.4967},
 primaryClass = {astro-ph.SR},
       adsurl = {https://ui.adsabs.harvard.edu/abs/2013ApJ...778..116N}
}

@ARTICLE{Park_2021ApJ...923..171P,
       author = {{Park}, Sunkyung and {K{\'o}sp{\'a}l}, {\'A}gnes and {Cruz-S{\'a}enz de Miera}, Fernando and {Siwak}, Micha{\l} and {Dr{\'o}{\.z}d{\.z}}, Marek and {Ign{\'a}cz}, Bernadett and {Jaffe}, Daniel T. and {K{\"o}nyves-T{\'o}th}, R{\'e}ka and {Kriskovics}, Levente and {Lee}, Jae-Joon and {Lee}, Jeong-Eun and {Mace}, Gregory N. and {Og{\l}oza}, Waldemar and {P{\'a}l}, Andr{\'a}s and {Potter}, Stephen B. and {Szab{\'o}}, Zs{\'o}fia Marianna and {Sefako}, Ramotholo and {Worters}, Hannah L.},
        title = "{V899 Mon: A Peculiar Eruptive Young Star Close to the End of Its Outburst}",
      journal = {\apj},
         year = 2021,
        month = dec,
       volume = {923},
       number = {2},
          eid = {171},
        pages = {171},
          doi = {10.3847/1538-4357/ac29c4},
archivePrefix = {arXiv},
       eprint = {2109.11283},
 primaryClass = {astro-ph.SR},
       adsurl = {https://ui.adsabs.harvard.edu/abs/2021ApJ...923..171P}
}

@ARTICLE{2011ApJ...736...72K,
       author = {{K{\'o}sp{\'a}l}, {\'A}. and {{\'A}brah{\'a}m}, P. and {Goto}, M. and {Reg{\'a}ly}, Zs. and {Dullemond}, C.~P. and {Henning}, Th. and {Juh{\'a}sz}, A. and {Sicilia-Aguilar}, A. and {van den Ancker}, M.},
        title = "{Near-infrared Spectroscopy of EX Lupi in Outburst}",
      journal = {\apj},
         year = 2011,
        month = jul,
       volume = {736},
       number = {1},
          eid = {72},
        pages = {72},
          doi = {10.1088/0004-637X/736/1/72},
archivePrefix = {arXiv},
       eprint = {1105.1287},
 primaryClass = {astro-ph.SR},
       adsurl = {https://ui.adsabs.harvard.edu/abs/2011ApJ...736...72K}
}

@ARTICLE{Banzatti_2019ApJ...870...76B,
       author = {{Banzatti}, Andrea and {Pascucci}, Ilaria and {Edwards}, Suzan and {Fang}, Min and {Gorti}, Uma and {Flock}, Mario},
        title = "{Kinematic Links and the Coevolution of MHD Winds, Jets, and Inner Disks from a High-resolution Optical [O I] Survey}",
      journal = {\apj},
         year = 2019,
        month = jan,
       volume = {870},
       number = {2},
          eid = {76},
        pages = {76},
          doi = {10.3847/1538-4357/aaf1aa},
archivePrefix = {arXiv},
       eprint = {1811.06544},
 primaryClass = {astro-ph.HE},
       adsurl = {https://ui.adsabs.harvard.edu/abs/2019ApJ...870...76B}
}

@ARTICLE{Beristain_2001ApJ...551.1037B,
       author = {{Beristain}, Georgina and {Edwards}, Suzan and {Kwan}, John},
        title = "{Helium Emission from Classical T Tauri Stars: Dual Origin in Magnetospheric Infall and Hot Wind}",
      journal = {\apj},
         year = 2001,
        month = apr,
       volume = {551},
       number = {2},
        pages = {1037-1064},
          doi = {10.1086/320233},
       adsurl = {https://ui.adsabs.harvard.edu/abs/2001ApJ...551.1037B}
}

@ARTICLE{Kwan_2011MNRAS.411.2383K,
       author = {{Kwan}, John and {Fischer}, William},
        title = "{Origins of the H, He I and Ca II line emission in classical T Tauri stars}",
      journal = {\mnras},
         year = 2011,
        month = mar,
       volume = {411},
       number = {4},
        pages = {2383-2425},
          doi = {10.1111/j.1365-2966.2010.17863.x},
archivePrefix = {arXiv},
       eprint = {1010.3265},
 primaryClass = {astro-ph.SR},
       adsurl = {https://ui.adsabs.harvard.edu/abs/2011MNRAS.411.2383K}
}

@ARTICLE{Nowacki_2023A&A...678A..86N,
       author = {{Nowacki}, H. and {Alecian}, E. and {Perraut}, K. and {Zaire}, B. and {Folsom}, C.~P. and {Pouilly}, K. and {Bouvier}, J. and {Manick}, R. and {Pantolmos}, G. and {Sousa}, A.~P. and {Dougados}, C. and {Hussain}, G.~A.~J. and {Alencar}, S.~H.~P. and {Le Bouquin}, J.~B.},
        title = "{Star-disk interactions in the strongly accreting T Tauri star S CrA N}",
      journal = {\aap},
         year = 2023,
        month = oct,
       volume = {678},
          eid = {A86},
        pages = {A86},
          doi = {10.1051/0004-6361/202347145},
archivePrefix = {arXiv},
       eprint = {2308.03442},
 primaryClass = {astro-ph.SR},
       adsurl = {https://ui.adsabs.harvard.edu/abs/2023A&A...678A..86N}
}

@INPROCEEDINGS{Lada_1987IAUS..115....1L,
       author = {{Lada}, Charles J.},
        title = "{Star formation: from OB associations to protostars.}",
    booktitle = {Star Forming Regions},
         year = 1987,
       editor = {{Peimbert}, Manuel and {Jugaku}, Jun},
       series = {IAU Symposium},
       volume = {115},
        month = jan,
        pages = {1},
       adsurl = {https://ui.adsabs.harvard.edu/abs/1987IAUS..115....1L}
}

@ARTICLE{Ellerbroek_2013A&A...551A...5E,
       author = {{Ellerbroek}, L.~E. and {Podio}, L. and {Kaper}, L. and {Sana}, H. and {Huppenkothen}, D. and {de Koter}, A. and {Monaco}, L.},
        title = "{The outflow history of two Herbig-Haro jets in RCW 36: HH 1042 and HH 1043}",
      journal = {\aap},
         year = 2013,
        month = mar,
       volume = {551},
          eid = {A5},
        pages = {A5},
          doi = {10.1051/0004-6361/201220635},
archivePrefix = {arXiv},
       eprint = {1212.4144},
 primaryClass = {astro-ph.SR},
       adsurl = {https://ui.adsabs.harvard.edu/abs/2013A&A...551A...5E}
}

@BOOK{2005oasp.book.....G,
       author = {{Gray}, David F.},
        title = "{The Observation and Analysis of Stellar Photospheres}",
         year = 2005,
          doi = {10.1017/CBO9781316036570},
       adsurl = {https://ui.adsabs.harvard.edu/abs/2005oasp.book.....G}
}

@ARTICLE{Calvet_1993ApJ...402..623C,
       author = {{Calvet}, Nuria and {Hartmann}, Lee and {Kenyon}, Scott J.},
        title = "{Mass Loss from Pre--Main-Sequence Accretion Disks. I. The Accelerating Wind of FU Orionis}",
      journal = {\apj},
         year = 1993,
        month = jan,
       volume = {402},
        pages = {623},
          doi = {10.1086/172164},
       adsurl = {https://ui.adsabs.harvard.edu/abs/1993ApJ...402..623C}
}

@INPROCEEDINGS{Bouvier_2007prpl.conf..479B,
       author = {{Bouvier}, J. and {Alencar}, S.~H.~P. and {Harries}, T.~J. and {Johns-Krull}, C.~M. and {Romanova}, M.~M.},
        title = "{Magnetospheric Accretion in Classical T Tauri Stars}",
    booktitle = {Protostars and Planets V},
         year = 2007,
       editor = {{Reipurth}, Bo and {Jewitt}, David and {Keil}, Klaus},
        month = jan,
        pages = {479},
          doi = {10.48550/arXiv.astro-ph/0603498},
archivePrefix = {arXiv},
       eprint = {astro-ph/0603498},
 primaryClass = {astro-ph},
       adsurl = {https://ui.adsabs.harvard.edu/abs/2007prpl.conf..479B}
}

@ARTICLE{Bailer_2021AJ....161..147B,
       author = {{Bailer-Jones}, C.~A.~L. and {Rybizki}, J. and {Fouesneau}, M. and {Demleitner}, M. and {Andrae}, R.},
        title = "{Estimating Distances from Parallaxes. V. Geometric and Photogeometric Distances to 1.47 Billion Stars in Gaia Early Data Release 3}",
      journal = {\aj},
         year = 2021,
        month = mar,
       volume = {161},
       number = {3},
          eid = {147},
        pages = {147},
          doi = {10.3847/1538-3881/abd806},
archivePrefix = {arXiv},
       eprint = {2012.05220},
 primaryClass = {astro-ph.SR},
       adsurl = {https://ui.adsabs.harvard.edu/abs/2021AJ....161..147B}
}

@ARTICLE{2021MNRAS.503..938Z,
       author = {{{\v{Z}}erjal}, Maru{\v{s}}a and {Rains}, Adam D. and {Ireland}, Michael J. and {Zhou}, George and {Kammerer}, Jens and {Wallace}, Alex and {Orenstein}, Brendan J. and {Nordlander}, Thomas and {Abbot}, Harrison and {Chang}, Seo-Won},
        title = "{A spectroscopically confirmed Gaia-selected sample of 318 new young stars within {\ensuremath{\sim}}200 pc}",
      journal = {\mnras},
         year = 2021,
        month = may,
       volume = {503},
       number = {1},
        pages = {938-952},
          doi = {10.1093/mnras/stab513},
archivePrefix = {arXiv},
       eprint = {2012.08756},
 primaryClass = {astro-ph.SR},
       adsurl = {https://ui.adsabs.harvard.edu/abs/2021MNRAS.503..938Z}
}

@ARTICLE{Campbell_2023A&A...673A..80C,
       author = {{Campbell-White}, Justyn and {Manara}, Carlo F. and {Sicilia-Aguilar}, Aurora and {Frasca}, Antonio and {Nielsen}, Louise D. and {Christian Schneider}, P. and {Nisini}, Brunella and {Bayo}, Amelia and {Ercolano}, Barbara and {{\'A}brah{\'a}m}, P{\'e}ter and {Claes}, Rik and {Fang}, Min and {Fedele}, Davide and {Gameiro}, Jorge Filipe and {Gangi}, Manuele and {K{\'o}sp{\'a}l}, {\'A}gnes and {Mauc{\'o}}, Karina and {Petr-Gotzens}, Monika G. and {Rigliaco}, Elisabetta and {Robinson}, Connor and {Siwak}, Michal and {Tychoniec}, Lukasz and {Venuti}, Laura},
        title = "{Empirical determination of the lithium 6707.856 {\r{A}} wavelength in young stars}",
      journal = {\aap},
         year = 2023,
        month = may,
       volume = {673},
          eid = {A80},
        pages = {A80},
          doi = {10.1051/0004-6361/202245696},
archivePrefix = {arXiv},
       eprint = {2303.03843},
 primaryClass = {astro-ph.SR},
       adsurl = {https://ui.adsabs.harvard.edu/abs/2023A&A...673A..80C}
}

@ARTICLE{Pascucci_2015ApJ...814...14P,
       author = {{Pascucci}, I. and {Edwards}, S. and {Heyer}, M. and {Rigliaco}, E. and {Hillenbrand}, L. and {Gorti}, U. and {Hollenbach}, D. and {Simon}, M.~N.},
        title = "{Narrow Na and K Absorption Lines Toward T Tauri Stars: Tracing the Atomic Envelope of Molecular Clouds}",
      journal = {\apj},
         year = 2015,
        month = nov,
       volume = {814},
       number = {1},
          eid = {14},
        pages = {14},
          doi = {10.1088/0004-637X/814/1/14},
archivePrefix = {arXiv},
       eprint = {1510.02022},
 primaryClass = {astro-ph.EP},
       adsurl = {https://ui.adsabs.harvard.edu/abs/2015ApJ...814...14P}
}

@ARTICLE{Micolta_2024ApJ...976..251M,
       author = {{Micolta}, Marbely and {Calvet}, Nuria and {Thanathibodee}, Thanawuth and {Magris C.}, Gladis and {Manara}, Carlo F. and {Venuti}, Laura and {Alcal{\'a}}, Juan Manuel and {Herczeg}, Gregory J.},
        title = "{Using the Ca II Lines in T Tauri Stars to Infer the Abundance of Refractory Elements in the Innermost Disk Region}",
      journal = {\apj},
         year = 2024,
        month = dec,
       volume = {976},
       number = {2},
          eid = {251},
        pages = {251},
          doi = {10.3847/1538-4357/ad8884},
archivePrefix = {arXiv},
       eprint = {2410.17327},
 primaryClass = {astro-ph.SR},
       adsurl = {https://ui.adsabs.harvard.edu/abs/2024ApJ...976..251M}
}

@ARTICLE{2005ApJ...626..498M,
       author = {{Mohanty}, Subhanjoy and {Jayawardhana}, Ray and {Basri}, Gibor},
        title = "{The T Tauri Phase Down to Nearly Planetary Masses: Echelle Spectra of 82 Very Low Mass Stars and Brown Dwarfs}",
      journal = {\apj},
         year = 2005,
        month = jun,
       volume = {626},
       number = {1},
        pages = {498-522},
          doi = {10.1086/429794},
archivePrefix = {arXiv},
       eprint = {astro-ph/0502155},
 primaryClass = {astro-ph},
       adsurl = {https://ui.adsabs.harvard.edu/abs/2005ApJ...626..498M}
}

@ARTICLE{Kun_2003A&A...398..169K,
       author = {{Kun}, M. and {Wouterloot}, J.~G.~A. and {T{\'o}th}, L.~V.},
        title = "{Probing the structure of a birthplace of intermediate-mass stars: Ammonia cores in Lynds 1340}",
      journal = {\aap},
         year = 2003,
        month = jan,
       volume = {398},
        pages = {169-180},
          doi = {10.1051/0004-6361:20021640},
archivePrefix = {arXiv},
       eprint = {astro-ph/0211180},
 primaryClass = {astro-ph},
       adsurl = {https://ui.adsabs.harvard.edu/abs/2003A&A...398..169K}
}

@ARTICLE{2024A&A...690A.225A,
       author = {{Armeni}, A. and {Stelzer}, B. and {Frasca}, A. and {Manara}, C.~F. and {Walter}, F.~M. and {Alcal{\'a}}, J.~M. and {Schneider}, P.~C. and {Sicilia-Aguilar}, A. and {Campbell-White}, J. and {Fiorellino}, E. and {Gameiro}, J.~F. and {Gangi}, M.},
        title = "{Evidence for magnetic boundary layer accretion in RU Lup: A spectrophotometric analysis}",
      journal = {\aap},
         year = 2024,
        month = oct,
       volume = {690},
          eid = {A225},
        pages = {A225},
          doi = {10.1051/0004-6361/202451065},
archivePrefix = {arXiv},
       eprint = {2408.14996},
 primaryClass = {astro-ph.SR},
       adsurl = {https://ui.adsabs.harvard.edu/abs/2024A&A...690A.225A}
}

@ARTICLE{Bouvier_2020A&A...643A..99B,
       author = {{Bouvier}, J. and {Alecian}, E. and {Alencar}, S.~H.~P. and {Sousa}, A. and {Donati}, J.-F. and {Perraut}, K. and {Bayo}, A. and {Rebull}, L.~M. and {Dougados}, C. and {Duvert}, G. and {Berger}, J.-P. and {Benisty}, M. and {Pouilly}, K. and {Folsom}, C. and {Moutou}, C. and {SPIRou Consortium}},
        title = "{Investigating the magnetospheric accretion process in the young pre-transitional disk system DoAr 44 (V2062 Oph). A multiwavelength interferometric, spectropolarimetric, and photometric observing campaign}",
      journal = {\aap},
         year = 2020,
        month = nov,
       volume = {643},
          eid = {A99},
        pages = {A99},
          doi = {10.1051/0004-6361/202038892},
archivePrefix = {arXiv},
       eprint = {2010.00887},
 primaryClass = {astro-ph.SR},
       adsurl = {https://ui.adsabs.harvard.edu/abs/2020A&A...643A..99B}
}

@ARTICLE{Petrov_2011A&A...535A...6P,
       author = {{Petrov}, P.~P. and {Gahm}, G.~F. and {Stempels}, H.~C. and {Walter}, F.~M. and {Artemenko}, S.~A.},
        title = "{Accretion-powered chromospheres in classical T Tauri stars}",
      journal = {\aap},
         year = 2011,
        month = nov,
       volume = {535},
          eid = {A6},
        pages = {A6},
          doi = {10.1051/0004-6361/201116721},
archivePrefix = {arXiv},
       eprint = {1109.1266},
 primaryClass = {astro-ph.SR},
       adsurl = {https://ui.adsabs.harvard.edu/abs/2011A&A...535A...6P}
}

@ARTICLE{Beristain_1998ApJ...499..828B,
       author = {{Beristain}, Georgina and {Edwards}, Suzan and {Kwan}, John},
        title = "{Permitted Iron Emission Lines in the Classical T Tauri Star DR Tauri}",
      journal = {\apj},
         year = 1998,
        month = may,
       volume = {499},
       number = {2},
        pages = {828-852},
          doi = {10.1086/305675},
       adsurl = {https://ui.adsabs.harvard.edu/abs/1998ApJ...499..828B}
}

@ARTICLE{Giannini_2019A&A...631A..44G,
       author = {{Giannini}, T. and {Nisini}, B. and {Antoniucci}, S. and {Biazzo}, K. and {Alcal{\'a}}, J. and {Bacciotti}, F. and {Fedele}, D. and {Frasca}, A. and {Harutyunyan}, A. and {Munari}, U. and {Rigliaco}, E. and {Vitali}, F.},
        title = "{GIARPS High-resolution Observations of T Tauri stars (GHOsT). I. Jet line emission}",
      journal = {\aap},
         year = 2019,
        month = nov,
       volume = {631},
          eid = {A44},
        pages = {A44},
          doi = {10.1051/0004-6361/201936085},
archivePrefix = {arXiv},
       eprint = {1909.10392},
 primaryClass = {astro-ph.SR},
       adsurl = {https://ui.adsabs.harvard.edu/abs/2019A&A...631A..44G}
}

@ARTICLE{Oh_2015JKAS...48..113O,
       author = {{Oh}, Heeyoung and {Pyo}, Tae-Soo and {Yuk}, In-Soo and {Park}, Byeong-Gon},
        title = "{Long-Slit Spectroscopy of Parsec-Scale Jets from DG Tauri}",
      journal = {Journal of Korean Astronomical Society},
         year = 2015,
        month = apr,
       volume = {48},
       number = {2},
        pages = {113-123},
          doi = {10.5303/JKAS.2015.48.2.113},
archivePrefix = {arXiv},
       eprint = {1505.00942},
 primaryClass = {astro-ph.SR},
       adsurl = {https://ui.adsabs.harvard.edu/abs/2015JKAS...48..113O}
}

@ARTICLE{Castro_2024A&A...688A...1C,
       author = {{Castro-Ginard}, Alfred and {Penoyre}, Zephyr and {Casey}, Andrew R. and {Brown}, Anthony G.~A. and {Belokurov}, Vasily and {Cantat-Gaudin}, Tristan and {Drimmel}, Ronald and {Fouesneau}, Morgan and {Khanna}, Shourya and {Kurbatov}, Evgeny P. and {Price-Whelan}, Adrian M. and {Rix}, Hans-Walter and {Smart}, Richard L.},
        title = "{Gaia DR3 detectability of unresolved binary systems}",
      journal = {\aap},
         year = 2024,
        month = aug,
       volume = {688},
          eid = {A1},
        pages = {A1},
          doi = {10.1051/0004-6361/202450172},
archivePrefix = {arXiv},
       eprint = {2404.14127},
 primaryClass = {astro-ph.GA},
       adsurl = {https://ui.adsabs.harvard.edu/abs/2024A&A...688A...1C}
}

@ARTICLE{Palla_1993ApJ...418..414P,
       author = {{Palla}, Francesco and {Stahler}, Steven W.},
        title = "{The Pre-Main-Sequence Evolution of Intermediate-Mass Stars}",
      journal = {\apj},
         year = 1993,
        month = nov,
       volume = {418},
        pages = {414},
          doi = {10.1086/173402},
       adsurl = {https://ui.adsabs.harvard.edu/abs/1993ApJ...418..414P}
}

@ARTICLE{2025A&A...700A..32R,
       author = {{Rota}, A.~A. and {van der Marel}, N. and {Garufi}, A. and {Carrasco-Gonz{\'a}lez}, C. and {Macias}, E. and {Pascucci}, I. and {Sellek}, A. and {Testi}, L. and {Isella}, A. and {Facchini}, S.},
        title = "{A correlation between accretion and outflow rates for class II young stellar objects with full and transition disks}",
      journal = {\aap},
         year = 2025,
        month = aug,
       volume = {700},
          eid = {A32},
        pages = {A32},
          doi = {10.1051/0004-6361/202554259},
archivePrefix = {arXiv},
       eprint = {2505.16586},
 primaryClass = {astro-ph.EP},
       adsurl = {https://ui.adsabs.harvard.edu/abs/2025A&A...700A..32R}
}

@ARTICLE{Simon_2016ApJ...831..169S,
       author = {{Simon}, M.~N. and {Pascucci}, I. and {Edwards}, S. and {Feng}, W. and {Gorti}, U. and {Hollenbach}, D. and {Rigliaco}, E. and {Keane}, J.~T.},
        title = "{Tracing Slow Winds from T Tauri Stars via Low-velocity Forbidden Line Emission}",
      journal = {\apj},
         year = 2016,
        month = nov,
       volume = {831},
       number = {2},
          eid = {169},
        pages = {169},
          doi = {10.3847/0004-637X/831/2/169},
archivePrefix = {arXiv},
       eprint = {1608.06992},
 primaryClass = {astro-ph.SR},
       adsurl = {https://ui.adsabs.harvard.edu/abs/2016ApJ...831..169S}
}
%do not change the name of bib file, it only takes 'thebibliography.bib' name as bib file

\end{document}